\documentclass[12pt]{article}
\pdfoutput=1

\usepackage{draft,hyperref,tikz,cancel,subfig,float,multirow,bbm,diagbox,hhline}

\usepackage{graphicx,etoolbox}
\robustify{\includegraphics}

\usepackage[nosort]{cite}

\usetikzlibrary{snakes}
\usetikzlibrary{shapes.misc}

\newsavebox{\ns}
\newsavebox{\dbrane}
\newsavebox{\dbshort}

\def\be{\begin{equation}}
\def\ee{\end{equation}}
\def\bea{\begin{eqnarray}}
\def\eea{\end{eqnarray}}

\newcommand\diff{\mathrm{d}}

\newcommand\Ie{\textit{i.e.}}

\newcommand{\ii}{\mathrm{i}}

\newcommand{\Real}{\mathrm{Re}\, }

\newlength{\sswidth}

\usepackage{color}
\usepackage{latexsym}
\usepackage{epsfig,amssymb,euscript, mathrsfs,cite}
\usepackage{arydshln} 
\usepackage{amsmath}
\definecolor{MyDarkBlue}{rgb}{0.15,0.15,0.45}

\begin{document}

\begin{titlepage}

\preprint{ CALT-TH 2017-058  \\ PUPT-2545}

\begin{center}

\hfill \\
\hfill \\
\vskip 1cm

\title{
Romans Supergravity from Five-Dimensional Holograms
}

\author{Chi-Ming Chang$^a$, Martin Fluder$^b$, Ying-Hsuan Lin$^b$, Yifan Wang$^c$
}

\address{${}^a$Center for Quantum Mathematics and Physics (QMAP) \\
University of California, Davis, CA 95616, USA}
\address{${}^b$Walter Burke Institute for Theoretical Physics \\ California Institute of Technology,
Pasadena, CA 91125, USA}
\address{${}^c$Joseph Henry Laboratories\\ Princeton University, Princeton, NJ 08544, USA}

\email{wychang@ucdavis.edu, fluder@caltech.edu, yhlin@caltech.edu, yifanw@princeton.edu}

\end{center}

\vfill

\abstract{We study five-dimensional superconformal field theories and their holographic dual, matter-coupled Romans supergravity.  On the one hand, some recently derived formulae allow us to extract the central charges from deformations of the supersymmetric five-sphere partition function, whose large $N$ expansion can be computed using matrix model techniques. On the other hand, the conformal and flavor central charges can be extracted from the six-dimensional supergravity action, by carefully analyzing its embedding into type I' string theory. The results match on the two sides of the holographic duality. Our results also provide analytic evidence for the symmetry enhancement in five-dimensional superconformal field theories.
}

\vfill

\end{titlepage}

\eject

\tableofcontents

\section{Introduction}

Since the discovery of AdS/CFT, the holographic equivalence between quantum gravity in anti-de Sitter space and conformal field theory has elucidated key aspects on both sides of the duality~\cite{Maldacena:1997re}.  Typically, strong coupling properties on one side are reincarnated with a simple weak coupling description on the other side, where physical observables can be computed in perturbation theory.  However, there are also holographic dualities that do not have genuine weak coupling corners in their parameter spaces.  Such is the case for interacting superconformal field theories in five dimensions, which do not admit marginal deformations, and have no known weak coupling limit~\cite{Cordova:2016xhm}. On the gravity side, the dilaton coupling of type I' string theory diverges towards the boundary of the internal space~\cite{Polchinski:1995df}.  Therefore, quantitative predictions of these holographic dualities are more difficult to come about.

First steps towards closing this gap have been taken in~\cite{Ferrara:1998gv,Brandhuber:1999np,Bergman:2012kr,Jafferis:2012iv,Bergman:2012qh,Assel:2012nf,Bergman:2013koa,Alday:2014rxa,Alday:2014bta,Alday:2014fsa,Hama:2014iea,Alday:2015jsa}.  On the gravity side, it has been realized that in an appropriate low energy regime, the type I' string theory admits an effective supergravity description, which upon reduction to six dimensions becomes Romans $F(4)$ supergravity~\cite{Romans:1985tw,Cvetic:1999un}  plus additional matter\cite{DAuria:2000afl,Andrianopoli:2001rs}.\footnote{Romans $F(4)$ (gauged) supergravity contains only the fields dual to the  five-dimensional  stress tensor multiplet in the boundary conformal field theory. It is a consistent truncation of the type I' reduction. Note that while the massive IIA supergravity was also discovered by Romans, we reserve the name of Romans supergravity for six dimensions.
}
It is expected that this much simpler six-dimensional effective description captures key aspects of five-dimensional superconformal field theories.  On the field theory side, following the seminal work of~\cite{Pestun:2007rz}, new advances in supersymmetric localization have provided extremely powerful tools for extracting supersymmetric observables even in these strongly interacting theories.  Careful analyses of the localized formulae for these observables in the large $N$ limit have provided checks of these holographic correspondences beyond kinematics~\cite{Jafferis:2012iv,Assel:2012nf,Alday:2014rxa,Alday:2014bta,Alday:2014fsa,Hama:2014iea,Alday:2015jsa}.

The present paper aims at a more thorough study of an enlarged set of observables in five-dimensional superconformal field theories, and their manifestations in the bulk supergravity.  The grander goal is to gain a clearer understanding of the strong coupling phenomena on both sides of the holographic correspondence, and to elucidate the general relationships connecting the rich landscape~\cite{Seiberg:1996bd,Morrison:1996xf,Intriligator:1997pq,Jefferson:2017ahm} of five-dimensional superconformal field theories, such as possible $F$- or $C$-theorems concerning renormalization group flows~\cite{Jafferis:2011zi,Klebanov:2011gs,Casini:2012ei} (see also~\cite{Pufu:2016zxm} and further references therein).  Many of these theories exhibit flavor symmetry enhancement, where extra conserved currents carrying instanton charges emerge when reaching the ultraviolet fixed point, and enhance the flavor symmetry~\cite{Seiberg:1996bd,Morrison:1996xf,Intriligator:1997pq}.  In the bulk, this enhancement is a non-perturbative phenomenon that involves D0-branes approaching, and D4-branes extended towards the strong coupling boundary where the dilaton diverges~\cite{Polchinski:1995df}.  Thus, a deep understanding of flavor symmetry enhancement may shed light on the non-perturbative dynamics in string theory.

The physical observables that we shall pursue are the supersymmetric (squashed) five-sphere partition function, the conformal central charge, and the various flavor central charges.  These charges appear in the superconformal block decomposition of the BPS four-point functions, and serve as inputs to the bootstrap analyses of these superconformal field theories~\cite{Chang:2017cdx}.  The flavor central charges also serve as indicators of symmetry enhancement, including the aforementioned one. They also signify special features of the particular holographically dual pair. For instance,
 we shall see that the flavor central charges for the mesonic ${\rm U}(1)_{\rm M}$ symmetry and the ${\rm SU}(2)_{\rm R}$ R-symmetry have a simple relation, which is {\it a priori} obscure from the field theory description, but clear in the dual supergravity\cite{Bergman:2012kr}.

In previous work~\cite{Chang:2017cdx}, the present authors used conformal perturbation theory to establish a precise relation between the central charges and the squashed five-sphere partition function with mass deformations, and computed the latter via supersymmetric localization.  The resulting central charges for the rank-one Seiberg theories led to a bootstrap analysis that revealed information about the non-BPS spectra in these theories.  We further found that the quantities we computed have surprisingly small instanton corrections, and the comparison of flavor central charges provided strong numerical evidence for flavor symmetry enhancement.

The present paper concerns the large $N$ regime, in the hope to understand the holographic duality for five-dimensional superconformal field theories, and to what extent Romans $F(4)$ supergravity  (plus additional matter) captures the hologram.  In addition to Seiberg theories, we extend our analysis to a larger class of orbifold theories  proposed by~\cite{Bergman:2012kr} .  Let us first review prior progress and results in this direction.  By localization, the perturbative (squashed) five-sphere partition function can be reduced to an $N$-dimensional Coulomb branch integral~\cite{Hosomichi:2012ek,Kallen:2012va,Kim:2012ava,Imamura:2012xg,Lockhart:2012vp,Imamura:2012bm}, which has been computed analytically in the large $N$ limit using matrix model techniques~\cite{Jafferis:2012iv,Alday:2014rxa}. It has further been argued that in this limit, the instanton contributions are exponentially suppressed, and thus, the perturbative results are exact~\cite{Jafferis:2012iv}. The matrix model of the squashed five-sphere partition function was studied in~\cite{Alday:2014rxa}, and the result was matched  holographically  with the properly renormalized on-shell action of Romans $F(4)$ supergravity in the bulk.

The progress we make in the present paper is as follows.
\begin{enumerate}
\item  Borrowing the results on  the matrix model for  the large $N$ squashed five-sphere partition function from~\cite{Alday:2014rxa}, we compute the conformal central charge  of the Seiberg theories in the large $N$ limit.
\item
We further study the matrix model for the mass-deformed five-sphere partition function.  In particular, we find that the round sphere free energy, the conformal central charge, and the mesonic and baryonic flavor central charges scale to leading order as $N^{5/2}$.  Similarly, the hypermultiplet and instantonic flavor central charges are found to scale as $N^{3/2}$.  The coefficients of the latter two exactly agree in the large $N$ limit, providing analytic evidence for flavor symmetry enhancement  in the Seiberg theories .
\item  A subtlety in the above analysis is the potential presence of Chern-Simons-like counter-terms.  We explicitly determine a scheme under which the one-point function of the instanton number current vanishes in the ultraviolet, as is required by conformal symmetry.
\item  Except the instantonic flavor central charge, all the other central charges in the field theory exactly agree with the couplings in   Romans $F(4)$ supergravity (coupled to additional vectors) obtained by a careful reduction from type I' supergravity.\footnote{We do not calculate the instantonic flavor central charge from supergravity in this paper due to the subtleties explained in Section~\ref{sect:D0CJ}, the resolution of which is left to future work.
}
This match provides further evidence of the suppression of instanton contributions at large $N$ .
\item  Finally, we generalize the above considerations to a larger class of orbifold theories~\cite{Bergman:2012kr}. We match not only the conformal, hypermultiplet, and mesonic flavor central charges,  but also a set of baryonic flavor central charges (up to an overall constant), on the two sides of the holographic duality.
\end{enumerate}

The rest of this paper is organized as follows.  In Section~\ref{Sec:Review}, we review our previous work~\cite{Chang:2017cdx}, and highlight the main ingredients that are relevant to the present analysis.  In Section~\ref{Sec:LargeN}, we employ supersymmetric localization and matrix model techniques to the Seiberg theories, to compute the   free energy and central charges in the large $N$ limit.  The relevant properties of the triple sine function and some finite $N$ numerics are provided in Appendices~\ref{App:triplesine} and~\ref{App:Numerics}.  In Section~\ref{Sec:Holography}, we look at the dual type I' string theory, examine its reduction to  matter-coupled  Romans $F(4)$ supergravity  with additional vectors , and thereby compute the free energy and central charges.  In Section~\ref{Sec:QuiverTh}, a similar analysis is performed for a more general class of orbifold theories.  Section~\ref{Sec:Conclusions} ends with concluding remarks and future directions.

\section{Review of key ingredients}
\label{Sec:Review}

This section presents a review of the key ingredients for computing the central charges in interacting five-dimensional superconformal field theories. First, we present the general definition of the (squashed) supersymmetric five-sphere partition function, the key localization results, and its relation to the conformal central charge and flavor central charges. We then discuss the admissible counter-terms that can appear (on a five-sphere background). Finally, we introduce the theories of primary interest -- the Seiberg theories. For the derivation and in-depth discussions, we refer the reader to an earlier paper by the present authors~\cite{Chang:2017cdx}.

\subsection{Supersymmetric five-sphere partition function}
\label{Sec:partfunc}

A large class of five-dimensional superconformal field theories has an infrared gauge theory phase.  This infrared Lagrangian description allows localization computations, the strong coupling limit of which recovers quantities at the ultraviolet fixed point.  This section is devoted to a review of the localization results of the five-sphere partition function for general gauge theories.

The present discussion largely omits instanton contributions, which are generally crucial for the consistency ({\it e.g.}, symmetry enhancement) of the partition function.  Nonetheless, as we shall argue in~Section~\ref{Sec:LargeN} (see also~\cite{Jafferis:2012iv}), in the large $N$ limit which could be compared to the corresponding (weakly coupled) supergravity dual, the instanton contributions are exponentially suppressed. Consequently, we may simply deal with the perturbative part.

The perturbative part of the squashed five-sphere partition function (\Ie~without instantons) has been computed in~\cite{Kallen:2012va,Kim:2012ava,Imamura:2012bm}.  More precisely, the localization formula was derived for squashed supersymmetric backgrounds that retain ${\rm U}(1) \times {\rm SU}(3) \subset {\rm SO}(6)$ isometry, and then conjectured for the most generic squashing with ${\rm U}(1)\times {\rm U}(1) \times {\rm U}(1)$ isometry. The metric of a generically squashed (unit) five-sphere is given by
\ie\label{eqn:sqmetric}
\diff s^{2} \ = \ \sum_{i=1}^{3} (\diff y_{i}^{2} + y_{i}^{2} \diff \phi_{i}^{2}) + \widetilde \kappa^{2} \left( \sum_{j=1}^{3} a_{j} y_{j}^{2} \diff \phi_{j} \right)^{2} \,, \quad \widetilde\kappa^{2} \ = \ \frac{1}{1- \sum_{j=1}^{3} y_{j}^{2} a_{j}^{2}} \,,
\fe
where $\omega_{j} = 1 + a_{j}$, $j=1,2,3$, are the real squashing parameters, $\phi_{j} \sim \phi_{j} + 2\pi$ are periodic coordinates, and $y_{i}$ are constrained such that $\sum_{j=1}^{3}y_{j}^{2} = 1$. The round sphere (in terms of polar coordinates) is given by setting $a_{j} = 0$. Thus, the latter part of the metric can be viewed as a perturbation of order $\cO(a_{j}^{2})$ to the round five-sphere metric.

For a general five-dimensional gauge theory with a simple rank-$N$ gauge group $G_{\rm g}$ and $N_{\rm f}$ hypermultiplets in the real representation $R_{f} \oplus \bar R_{f}$ of $G_{\rm g}$, $f = 1, \dotsc, N_{\rm f}$, the perturbative squashed five-sphere partition function is given by\footnote{We use $\lambda$ to denote collectively the Cartan generators, and
\ie
\lambda_i = \Tr(T^i\lambda), \quad \Tr(T^a T^b)=\delta_{ab}.
\fe
Each $\A$ in the root lattice is a linear map from the Cartan subgalgebra to real numbers, hence $\A(\lambda) \in \bR$.
}
\ie\label{Eqn:Zpert}
\cZ_{\rm pert} \ = \ \frac{S_{3}^{\prime}\left( 0 \mid \vec{\omega}  \right)^{N}}{\left| \cW \right|} \left( \prod_{i=1}^{N} \int_{-\infty}^{\infty} \frac{\diff \lambda_i}{2\pi} \right) \, e^{-\frac{(2\pi)^{3}}{\omega_1\omega_2\omega_3} \mathfrak{F}(\lambda)}
\frac{ \prod_{\alpha} S_{3}\left( -\ii \alpha(\lambda) \mid \vec{\omega}  \right)}{\prod_f \prod_{\rho_{f}}S_{3}\left( \ii \rho_{f}(\lambda)+ \tfrac{\omega_{\mathrm{tot}}}{2} \mid \vec{\omega} \right)}\,,
\fe
where the products are taken over all the roots $\alpha$ of $G_{\rm g}$, the flavors $f = 1, \dotsc, N_{\rm f}$ that run over the hypermultiplets on which the flavor symmetry group $G_{\rm f}$ acts, and the weights $\rho_f$ of the particular representations $R_{f} \oplus \bar R_{f}$. The (Roman font) subscripts g and f denote gauge and flavor, respectively. We introduced the classical (flat space) prepotential $\mathfrak{F}(\lambda)$, which is given by
\ie
\mathfrak{F}(\lambda)  \ = \ \frac{1}{2 g_{\rm YM}^{2}} \Tr \lambda^{2} + \frac{k}{6} \Tr \lambda^{3} \,,
\fe
with $g_{\rm YM}$ the classical gauge coupling, $k$ the Chern-Simons coupling, and $\Tr(\cdot)$ is the Killing form defined as ($h^\vee$ is the dual Coxeter number)
\ie\label{Killing}
\Tr(\cdot) \ \equiv \ {1\over 2h^\vee}\tr_{\bf adj}(\cdot) \, .
\fe
Finally, $S_{3}$ is the triple sine function, and 
\ie
S_{3}^{\prime} (0 \mid \vec{\omega}) \ = \ \lim_{x\to 0}S_{3} (x \mid \vec{\omega}) \,.
\fe
We refer the reader to Appendix~\ref{App:triplesine} for a definition of $S_3$, and some of its properties relevant for the present paper.

There is another type of deformations.  For a theory of given flavor symmetry group $G_{\rm f}$, we can introduce mass parameters into the partition function by coupling the hypermultiplets to background vector multiplets.  For an Hermitian mass matrix $\cM \in \mathfrak{g}_{\rm f} \equiv Lie (G_{\rm f})$, the mass term is explicitly given by
\ie\label{eqn:hyperMassAction}
\int \diff^5 x \sqrt{g} \Big(-\epsilon^{ij}\bar q_i \cM^2 q_j+2\ii  t^{ij}  \bar q_i  \cM q_j-2  \bar \psi \cM \psi\Big),
\fe
where $q_i$ and $\bar q_i$ are the scalars, and $\psi$ the fermion, in the hypermultiplet.\footnote{Although the physical mass matrix $\cM$ in~\eqref{eqn:hyperMassAction} is Hermitian, when performing localization, one has to analytically continue $\cM$ to an anti-Hermitian matrix for convergence~\cite{Hosomichi:2012ek,Kallen:2012va}.
}
These masses arise from turning on vacuum expectation values for the scalars in the {\it background} vector multiplets, akin to Coulomb branch masses from turning on scalars in dynamical vector multiplets.  Hence, these generic mass terms appear in the partition function~\eqref{Eqn:Zpert} in the same way as the Coulomb branch parameters. We defer the presentation of the explicit formula for ${\cal Z}_{\rm pert}$ for Seiberg theories with a particular choice of mass deformations to Section~\ref{Sec:En}.

\subsection{Central charges from deformations of five-sphere partition function}
\label{Sec:Centralcharges}

In~\cite{Chang:2017cdx}, the present authors derived formulae for five-dimensional superconformal field theories that relate the conformal central charge and flavor central charges to deformations of the five-sphere partition function. The proof proceeds by coupling the five-sphere background to the appropriate background supergravity multiplets. Here, we explain the rationale and present the resulting formulae.

\subsubsection{Conformal central charge from metric deformations}

In order to extract the conformal central charge $C_T$ from a partition function, we study the superconformal field theory on a five-sphere background perturbed by coupling the stress tensor multiplet to a background supergravity multiplet on a generically \emph{squashed} five-sphere with metric given in~\eqref{eqn:sqmetric}. In order to preserve the full superconformal symmetry, we couple the theory to the five-dimensional $\cN=1$ standard Weyl multiplet 
\ie
\left( ~ g_{\m\n}, ~~ D, ~~ V_{\m}^{ij}, ~~ v_{\m\n}, ~~ \psi_{\m}^{i}, ~~ \chi^{i} ~ \right) \,,
\fe
which consists of the dilaton $D$, the metric $g_{\mu\nu}$, an ${\rm SU}(2)_{\rm R}$ symmetry gauge field $V_{\mu}{}^{ij}$, a two-form field $v_{\mu\nu}$, together with their fermionic partners -- the gravitino $\psi_{\mu}^{i}$ and the dilatino $\chi^{i}$. We deform the five-sphere background by writing $g_{\mu\nu} = g^{{\rm S}^5}_{\mu\nu} + h_{\mu\nu}$.   Upon doing so, the (bosonic) linearized action is given by
\ie\label{eqn:linearizedCT}
\delta S \ = \ \int \diff^{5}x \, \sqrt{g} \left( -h^{\mu\nu} T_{\mu\nu} + 2 V_{\mu}^{ij} J^{\mu}_{ij} - \ii v_{\mu\nu} B^{\mu\nu} + \frac{1}{8} D \Phi\right) + \cO( h_{\mu\nu}^{2} ) \,,
\fe
where $T_{\m\n}$, $J^{\mu}_{ij}$, $B_{\m\n}$ and $\Phi$ are the (bosonic) components of the stress tensor multiplet.\footnote{Notice that in the ``rigid'' limit, all fermionic fields of the background supergravity multiplets are set to zero.
\label{Footnote:Rigid}
}
The linearized background fields of the standard Weyl multiplet for the generic squashed five-sphere can be explicitly determined by solving the relevant supersymmetry conditions for the given metric.

From the linearized perturbation~\eqref{eqn:linearizedCT}, it follows that the five-sphere free energy at second order in the squashing parameters $a_{j}$ can be written in terms of the two-point functions of the currents $B_{\mu\nu}$, $J_{\mu}{}^{ij}$ and $\Phi$ on the round sphere. These two-point functions are related to the conformal central charge upon stereographic projection from flat space. By evaluating them explicitly, and expanding the background values for the bosonic fields in the standard Weyl multiplet to leading order in the squashing parameters, we find that the free energy is in fact related to the conformal central charge $C_T$ as
\ie\label{CTRelation}
F\big|_{a_{i}^{2}} \ = \ - \frac{\pi^{2} C_{T}}{1920} \left( \sum_{i=1}^{3} a_{i}^{2} - \sum_{i<j} a_i a_j \right) \,.
\fe

\subsubsection{Flavor central charge from mass deformations}

The same overall logic applies to extracting the flavor central charge $C_{J}^{G_{\rm f}}$ for a given flavor symmetry group $G_{\rm f}$, by considering mass deformations of the five-sphere free energy. Working in conformal perturbation theory, we couple the flavor current multiplet in the five-dimensional superconformal field theory to a background vector multiplet 
\ie\label{eqn:bgvm}
\left( ~ W^{a}_{\mu}, ~~ M^{a}, ~~ \Omega^{a\,  i}, ~~ Y^{a\, ij} ~ \right),
\fe
which contains the vector field $W^{a}_{\mu}$, the scalar $M^{a}$, the gaugino $\Omega^{a \, i}$ and the triplet of auxiliary fields $Y^{a \, ij}$, all in the adjoint representation of $G_{\rm f}$. The linearized perturbation of the \emph{bosonic} action is then given by$^\text{\ref{Footnote:Rigid}}$
\ie
\delta S \ = \ \int \diff^{5} x \sqrt{g} \left( 2 Y^{aij} L_{ij}^{a} - W^{a}_{\mu} J^{a\mu} + M^{a}N^{a} \right) \,,
\fe
where $L_{ij}^{a}$, $J^{a\mu}$, and $N^{a}$ are the bosonic components of the flavor current multiplet. On the five-sphere, the vector multiplet has to satisfy supersymmetry conditions, which can be solved by $W_{\mu} = \Omega^{i} = 0$ and $Y^{ij} = - M t^{ij}$, for $t^{ij} = \frac{\ii}{2} \sigma_{3}$ and some $M \in \mathfrak{g}_{\rm f} \equiv Lie(G_{\rm f})$. Thus, the leading order piece in the mass deformation of the five-sphere free energy can be expressed in terms of the two-point functions of $L_{ij}^{a}$ and $N^{a}$ on the round sphere. Those two-point functions are related to the corresponding flavor central charge $C_{J}^{G_{\rm f}}$ in flat space by stereographic projection, and we end up with
\ie\label{CJRelation1}
 F\big|_{M^{2}} \ = \ \frac{3\pi^{2} C_{J}^{G_{\rm f}}}{256} \delta_{ab} M^{a} M^{b} \,.
\fe

It remains to translate the mass parameters $M^{a}$ appearing in conformal perturbation theory to the mass matrix appearing in the (infrared) gauge theory Lagrangian. We refer the reader to~\cite{Chang:2017cdx} for a in-depth treatment of this. Here, we shall simply state the result,
\ie\label{CJRelation2}
 F\big|_{\cM^{2}} \ = \ \frac{3\pi^{2} C_{J}^{K_{\rm f}}}{512 \, {I}_{R_{K_{\rm f}}}} \tr_{R_{K_{\rm f}}} (\cM^{2}) \,,
\fe
where $R_{K_{\rm f}}$ is the representation of the hypermultiplets under $K_{\rm f}$ -- the manifest flavor symmetry group of the infrared Lagrangian, ${I}_{R_{K_{\rm f}}}$ is the Dynkin index associated with the represenation $R_{K_{\rm f}}$, and $\cM$ is the mass matrix in the action deformation \eqref{eqn:hyperMassAction}.  

The global symmetry can be enhanced at the ultraviolet fixed point, which is in fact a common phenomenon in five dimensions. In that case, if $G$ is simple, one can obtain $C_{J}^{G_{\rm f}}$ from $C_{J}^{K_{\rm f}}$, by use of the embedding index $I_{\mathfrak{k}\hookrightarrow \mathfrak{g}} $, \Ie,
\ie
\label{CJEmbedding}
C_{J}^{K_{\rm f}} \ = \ I_{\mathfrak{k}_{\rm f} \hookrightarrow \mathfrak{g}_{\rm f}} \, C_{J}^{G_{\rm f}} \,.
\fe

\subsection{Chern-Simons-like counter-terms}
\label{Sec:counter-terms}

The five-sphere free energy contains ultraviolet divergences.  Therefore, in order to get a physically relevant quantity, one must remove such divergences by introducing local diffeomorphism-invariant counter-terms, which would generically break conformal invariance. For the case of five-dimensional $\cN=1$ superconformal field theories, such counter-terms would break the superconformal group to $\mathfrak{su}(4|1)$, but renders the finite part of the five-sphere partition function unambiguous and physical. 

More explicitly, the counter-terms are supersymmetric completions of (mixed) Chern-Simons terms in Poincar\'e supergravity, and have been classified in~\cite{Chang:2017cdx} by the present authors. The three possible counter-terms involving a single scalar $M$ in the background vector multiplet \eqref{eqn:bgvm} are given by
\ie
\cL_{TTJ}^{(1)} \ &\ni \  \ii \kappa_{TTJ}^{1} \Big[\sqrt{g} M R^{2}
+\cdots \Big] \,,\\
\cL_{TTJ}^{(2)} \ &\ni \  \ii \kappa_{TTJ}^{2} \Big[\sqrt{g} M \left( R_{\m\n}R^{\m\n} -\frac{1}{8}R^{2} \right)
+\cdots \Big] \,,\\
\cL_{TTJ}^{(3)} \ &\ni \  \ii \kappa_{TTJ}^{3} \Big[\sqrt{g} M C_{\m\n\rho\sigma}C^{{\m\n\rho\sigma}}
+\cdots \Big] \,,
\fe
where $R$, $R_{\m\n}$ are the Ricci scalar and tensor, and $C_{\m\n\rho\sigma}$ is the Weyl tensor. Unitarity of the theory restricts the constants $\kappa_{TTJ}^{j}$ to be real, and upon localization each of those terms will contribute a \emph{real} piece at linear order in the (real) parameters $m_f$, where $m_f$ are the demotion of $M_f$ to constants, interpreted as mass parameters for the flavor symmetry.\footnote{Recall that localization requires $M$ to be imaginary and here $M$ is proportional to $\ii m_f$ by a real number.} If more than one background vector multiplet are present, then there exists another counter-term given by 
\ie
\cL_{JJJ} \ &\ni \ \ii \kappa_{JJJ}^{abc} \Big[\sqrt{g} M^{a} M^{b} M^{c} R
+\cdots \Big] \,.
\fe
As before, the constants $\kappa_{JJJ}^{abc}$ are real, and we conclude that such a counter-term contributes a \emph{real} piece to the localized five-sphere partition function, at cubic order in the mass parameters $m_f$ for the flavor symmetry. As we shall see in Section~\ref{Sec:FlavorU1J}, one has to subtract such terms off the large $N$ free energy to end up with a physically consistent result.

\subsection{Seiberg theories}
\label{Sec:En}

In the next few sections, we focus on a particular class of five-dimensional superconformal field theories, which were first introduced by Seiberg~\cite{Seiberg:1996bd}.\footnote{In Section~\ref{Sec:QuiverTh}, we consider a more general class of theories labeled by $n\in \mathbb{Z}_{\geq 1}$, of which $n=1$ corresponds to the Seiberg theories.
}
They can be constructed in type IIA string theory (or rather type I' string theory) by a D4-D8/O8-brane setup. We begin with two orientifold O8-planes located at $x^{9}=0$ and $x^{9}=\pi$, and put $0\leq N_{\rm f}<8$ D8-branes coinciding with the O8 at $x^{9}=0$ and $16-N_{\rm f}$ coinciding with the O8 at $x^{9}=\pi$. Finally, we add $N$ D4-branes inside the D8/O8-branes. In the limit of infinite string coupling, the D0-branes become massless, and together with the massless string degrees of freedom localized on the D4-branes yield the aforementioned Seiberg theories. The flavor symmetries for the Seiberg theories of arbitrary rank are given by 
\ie
E_{N_{\rm f}+1} \times {\rm SU}(2)_{\rm M}\,,
\fe
 where the first factor is enhanced from the ${\rm SO}(2N_{\rm f})$ symmetry of the $N_{\rm f}$ D8-branes and the ${\rm U}(1)_{\rm I}$ instanton particle (D0-brane) symmetry\cite{Polchinski:1995df,Seiberg:1996bd}. The latter factor is the mesonic symmetry arising from rotations in the $\{x^5,x^6,x^7,x^8\}$ directions.

\begin{table}[h]
\centering
\begin{tabular}{|c|cccccccccc|} \hline
&0&1&2&3&4&5&6&7&8&9\\
\hline\hline
D8/O8 &$\times$&$\times$&$\times$&$\times$&$\times$&$\times$&$\times$&$\times$&$\times$&  \\
\hline
D4&$\times$&$\times$&$\times$&$\times$&$\times$&&&&&\\
\hline
\end{tabular}
\caption{The D4-D8/O8-brane system in type I' string theory.}
\label{table:BraneDirections}
\end{table}

A renormalization group flow connects a Seiberg theory of flavor symmetry $E_{N_{\rm f}+1}$ in the ultraviolet to an infrared ${\rm USp}(2N)$ gauge theory. The D4-D4 strings give rise to a vector multiplet and a hypermultiplet in the antisymmetric representation of ${\rm USp}(2N)$. The D4-D8 strings give rise to $N_{\rm f} \leq 7$ fundamental hypermultiplets. Starting from the ultraviolet, the infrared gauge theory can be reached by turning on a deformation that becomes the five-dimensional Yang-Mills term towards the end of the flow. Although an infinite number of irrelevant terms arise along this flow, they are believe to be $Q$-exact, and thus do not contribute to the localized path integral~\cite{Kim:2012ava,Jafferis:2012iv,Kim:2012qf}.  Therefore, the supersymmetric partition function of the ultraviolet superconformal Seiberg theory is fully captured by the infrared ${\rm USp}(2N)$ gauge theory.

The supersymmetric five-sphere partition function has been computed by localization using the infrared gauge theory description. The result is a sum over the contributions from infinitely many saddle points. We presently review the result of the perturbative saddle point, where all the hyper- and vector multiplets have trivial vacuum expectation value~\cite{Kallen:2012va,Kim:2012ava,Imamura:2012bm}. For simplicity, we choose the mass matrix \eqref{eqn:hyperMassAction} for the hypermultiplet ${\rm SO}(2N_{\rm f})$ flavor symmetry to be
\ie
 \cM({\rm SO}(2N_{\rm f})) \ = \ m_\mathrm{f} \,\left( \ii \sigma_2 \otimes \mathbbm{1}_{N_{\rm f}} \right)& \in &  \mathfrak{so}\left( 2N_{\rm f} \right) \,,
\fe
and for the mesonic ${\rm SU}(2)_{\rm M}$ symmetry acting on antisymmetric hypermultiplet to be
\ie
\cM ({\rm SU}(2)_{\rm M}) \ = \ \ii m_{\mathrm{as}} \sigma_{3} &\in & \mathfrak{su}\left( 2 \right)_{\rm M} \,.
\fe
The corresponding perturbative squashed five-sphere partition function for the infrared ${\rm USp}(2N)$ gauge theories with $N_{\rm f}$ fundamental and one antisymmetric hypermultiplets is given by\footnote{Here, we adopt the notation $S_3(\pm x \mid \vec\omega) \equiv S_3(x \mid \vec\omega) S_3(-x \mid \vec\omega)$.
}
\ie\label{eqn:USpPartition}
& \hspace{-.2in} \cZ^{\rm pert} \ = \ \frac{S_{3}^{\prime}\left( 0 \mid \vec{\omega}  \right)^{N}}{\left| \cW \right|S_3 \left(\ii m_{\rm as}+\tfrac{\omega_{\mathrm{tot}}}{2}\mid\vec{\omega} \right)^{N-1}}
\left[ \int_{-\infty}^{\infty} \prod_{i=1}^{N} \frac{\diff \lambda_i}{2\pi}  \right]
\Bigg[
\exp\left( {-\frac{1}{\omega_1\omega_2\omega_3}  \frac{4\pi^{3}}{g_{\rm YM}^{2}} \sum_{i=1}^{N} \lambda_i^{2}}  \right) \\
&
\times\frac{
\prod_{i> j} S_{3}\left( \ii [\pm\lambda_i\pm\lambda_j] \mid \vec{\omega}  \right) 
\prod_{i=1}^{N} S_{3}\left( \pm 2 \ii\lambda_i \mid \vec{\omega}  \right)}
{
\prod_{i> j} S_{3}\left( \ii [\pm\lambda_i \pm \lambda_j]  + \ii m_{\rm as} + \tfrac{\omega_{\mathrm{tot}}}{2} \mid \vec{\omega} \right)
\prod_{i=1}^{N} S_{3}\left(  \pm\ii\lambda_i + \ii m_{\rm f}+ \tfrac{\omega_{\mathrm{tot}}}{2} \mid \vec{\omega} \right)^{N_{\rm f}}
}
\Bigg]\,,
\fe
where $|{\cal W}|=2^N N!$ is the order of the Weyl group of ${\rm USp}(2N)$. Note that since the ${\rm USp}(2N)$ gauge group has no cubic Casimir, the (gauge) Chern-Simons terms are absent, and the prepotential only consists of the classical Yang-Mills piece.

The dependence of the supersymmetric five-sphere partition function on the squashing and mass parameters encodes the central charges of the superconformal field theory. The embedding indices, which by \eqref{CJEmbedding} relate the flavor central charges of the infrared global symmetry group to the ultraviolet enhanced global symmetry group, in the case of Seiberg theories, are
\ie
\label{Embedding}
I_{\mathfrak{so}(2N_{\rm f}) \hookrightarrow \mathfrak{e}_{N_{\rm f}+1}} \ = \ 1, \qquad I_{\mathfrak{u}(1)_{\rm I} \hookrightarrow \mathfrak{e}_{N_{\rm f}+1}} \ = \ {4 \over 8-N_{\rm f}}.
\fe
We refer to~\cite{Chang:2017cdx} for the explicit computations.

\section{The large $N$ limit of Seiberg theories}\label{Sec:LargeN}

Motivated by holography, we proceed to the computation of the deformed five-sphere partition function for the Seiberg theories in the large $N$ limit. As we shall we, the supersymmetric partition function, which is hard to evaluate at finite $N$, simplifies in this limit.

To describe the partition function at the superconformal fixed point, we are required to send $g_{\text{\tiny YM}}\to \infty$ under which \emph{all} saddles in the instanton expansion contribute equally (without manifest suppression by a small parameter), and a perturbative analysis seems to break down. Nonetheless, it has been argued that the instanton contributions are exponentially suppressed in the large $N$ limit~\cite{Jafferis:2012iv}. To understand this, we write the full partition function schematically as
\ie
\mathcal{Z} \ \sim \  \sum_{n=0}^{\infty} \mathcal{Z}^{n} q^{n} \,,
\fe
where $\mathcal{Z}^{n}$ are functions of the fugacities that are only explicitly known for small instanton numbers, and
\ie
q \ = \ \exp\left({-\frac{8\pi^3}{g_{\text{\tiny YM}}^{2}}}\right) \,,
\fe
with $g_{\rm YM}$ being the infrared gauge coupling. When moving out on the Coulomb branch, the effective Yang-Mills coupling receives a one-loop correction
\ie
\frac{1}{g_{\mathrm{eff}}^{2}} -\frac{1}{g_{\text{\tiny YM}}^{2}} \ \propto \  \left| \lambda \right| \,,
\fe
where by $\lambda$ we collectively denote the Coulomb branch parameters of the theory.  Since the instanton particles are BPS, their masses are determined by the central charge (of the supersymmetry algebra), which is a linear combination of the bare gauge coupling and the Coulomb branch parameters. Accordingly, the effective parameter for the instanton expansion is governed by not the bare gauge coupling but an effective coupling,
\ie
q_{\mathrm{eff}}  \ &= \   \exp\left({-\frac{8\pi^3}{g_{\mathrm{eff}}^{2}}}\right)\,.
\fe
As we shall see below, in the large $N$ limit, the Coulomb branch parameters $\lambda$ scale asymptotically as 
\ie
\lambda \ \sim \ N^{1/2} \,, \qquad N \ \rightarrow \ \infty \,.
\fe
Consequently, the contributions with nontrivial instanton numbers are exponentially suppressed in the large $N$ limit.

The upshot is that in the large $N$ limit, the perturbative part of (squashed) five-sphere partition function is \emph{exact}. We can therefore explicitly determine the conformal central charge and the flavor central charges for the $E_{N_{\rm f}+1} \times {\rm SU}(2)_{\rm M}$ flavor symmetry, via the perturbative localization formula \eqref{eqn:USpPartition}, as well as the prescription of Section~\ref{Sec:Centralcharges}. The enhancement of the flavor symmetry from the manifest infrared ${\rm SO}(2N_{\rm f}) \times {\rm U}(1)_{\rm I}$ to the ultraviolet $E_{N_{\rm f}+1}$ implies that the values of the exceptional flavor central charge computed from the ${\rm SO}(2N_{\rm f})$ and ${\rm U}(1)_{\rm I}$ embeddings should agree.  Whereas numerical evidence for the agreement was found in~\cite{Chang:2017cdx} for the rank-one Seiberg theories, here we shall find an \emph{exact} agreement in the large $N$ limit. In Section~\ref{Sec:Holography}, the results of the present section will be compared with the central charges extracted directly from the dual supergravity description.

\subsection{Free energy and conformal central charge}

Let us first recall the large $N$ computation of the  undeformed free energy in~\cite{Jafferis:2012iv,Alday:2014rxa,Alday:2014bta}. We start by rewriting the perturbative partition function \eqref{eqn:USpPartition} into the form
\ie
\mathcal{Z}^{\rm pert}  \  = \ \frac{1}{|\mathcal{W}|} \int  \left[ \diff \lambda \right]  \ e^{-F(\lambda)} \, . \label{spf}
\fe
where we have chosen to represent the exponent by the symbol $F(\lambda)$ because when later evaluated at the large $N$ saddle point, it becomes the large $N$ free energy $F$. In the following, $F(\lambda)$ will be referred to as the ``localized action''. For the Seiberg theories,\footnote{Here, we adopt the notation $G(\pm z) \equiv G(z) + G(-z)$. Our $G_V$ and $G_H$ are related to the $F_V$ and $F_H$ in~\cite{Jafferis:2012iv} by
\ie
F_V(z) \ = \ {1\over 2}G_V(\pm z\mid 1,1,1), \qquad F_H(z) \ = \ G_H( z\mid 1,1,1).
\fe
}
\ie
 F( \lambda ) \ = \ 
& 
\sum_{i=1}^{N} \left[ 
G_{V}(\pm2 \lambda_i\mid \vec{\omega}) + 
N_{\rm f} \, G_{H}(\pm\lambda_i\mid \vec{\omega}) \right]
\\
& + \sum_{i \neq j}^{N} \left[ {1\over2} 
 G_{V} (\pm \lambda_i \pm \lambda_j \mid \vec{\omega})+
 G_{H} (\lambda_i \pm \lambda_j \mid \vec{\omega}) \right]
\, ,
\label{freeenergy}
\fe
where $G_V$ and $G_H$ are the logarithms of the triple sine functions, \Ie,
\ie
G_V \left( z \mid \vec{\omega}\right) \ = \ & - \log S_3\left( \ii z \mid \vec{\omega} \right)\,,
\\
G_H \left( z \mid \vec{\omega}\right) \  = \ & \log S_3\left( \ii z + \tfrac{\omega_{\mathrm{tot}}}{2} \mid\vec{\omega}\right)\,,
\fe
with $\omega_\mathrm{tot} = \omega_1 + \omega_2 + \omega_3$. By Weyl reflections of the ${\rm USp}(2N)$ gauge group, we can restrict the integration region to $\lambda_i \geq 0$, and compensate by a symmetry factor of $2^N$.

We now proceed to performing the large $N$ saddle point approximation of the integral \eqref{spf}. These saddle points can be studied numerically at large but finite $N$, and the results suggest that the Coulomb branch parameters scale (to leading order) as $\lambda_i \ \sim \ N^{1/2} x_i$ in the asymptotic limit of large $N$ and finite $x$. This particular scaling of $\lambda_i$ with $N$ can be argued analytically, as follows. First, assume that
\ie
\lambda_i \ = \ N^{\alpha} x_i \,, \quad \text{with}  \quad \alpha > 0\,.
\fe
Then, note that the functions $G_{V} (\pm z \mid \vec{\omega}) $ and $G_H( z \mid \vec{\omega}) $ have the following asymptotic expansions at $\left| z \right| \rightarrow \infty$,
\ie
G_{V} (\pm z \mid \vec{\omega}) 
 & \ \sim \ \frac{\pi}{3 \, \omega_1\omega_2\omega_3} |z|^{3} - \frac{\pi \left( \omega_{\mathrm{tot}}^{2} + \omega_1 \omega_2+\omega_1 \omega_3+\omega_2 \omega_3 \right)}{6 \,\omega_1\omega_2\omega_3} |z| \, ,
 \\
G_{H} ( z \mid \vec{\omega}) 
& \ \sim \  - \frac{\pi}{6 \, \omega_1\omega_2\omega_3} |z|^{3} - \frac{\pi \left( \omega^{2}_1+\omega^{2}_2 +\omega^{2}_3 \right)}{24 \, \omega_1\omega_2\omega_3} |z|\,,\label{expansionhyper}
\fe
which follow from the asymptotic formulae \eqref{S3Asym1} and \eqref{S3Asym2} for triple sine functions. As reviewed in Appendix~\ref{AsymptTripSine}, the expansions \eqref{expansionhyper} have no subleading power law correction, a fact that will be important later.  The leading order term in the first line of~\eqref{freeenergy} scales as $N^{1+3 \alpha}$, whereas the leading order term in the second line of~\eqref{freeenergy} scales as $N^{2 + \alpha}$. In order to get a nontrivial saddle point, both terms must contribute to the same order, and we thereby determine $\alpha = {1\over2}$.  

In the large $N$ limit, we introduce a density $\rho(x)$ for the rescaled Coulomb branch parameters $x_i$ (we use the Weyl reflections of the ${\rm USp}(2N)$ gauge group to restrict to $x_i \geq 0$),
\ie
\rho (x) \ \equiv \ \frac{1}{N} \sum^{N}_{i=1} \delta \left( x-x_i \right) \, ,
\fe
normalized such that
\ie
\int \diff x \, \rho(x) \ = \ 1 \,. \label{eqn:normalizRho}
\fe
In the continuum limit, the localized action \eqref{freeenergy} becomes
\ie\label{Eqn:Fasymsq}
\hspace{-.15in}
 F[\rho] \ &= \ - {N^{5/2} \over \omega_1 \omega_2 \omega_3} \int_{0}^{x_\star} \diff x \, \rho(x) \int_{0}^{x_\star} \diff y \, \rho(y)  \left[ \frac{\pi \omega_{\mathrm{tot}}^{2}}{8} \left( x+y+|x-y| \right)  - \frac{(8-N_{\rm f}) \pi}{3} x^{3}\right]
 \\
& \hspace{1.5in} 
+ {\cal O}( N^{3/2} ) \, .
\fe
We are thus left with the simple variational problem of finding the function $\rho (x)$ that extremizes $F(\rho)$.  We add a Lagrange multiplier term for the constraint~\eqref{eqn:normalizRho},
\ie
\mu \left( \int_0^{x_\star} \diff  x \, \rho(x) - 1 \right),
\fe
and solve ${\D F}/{\D \rho} = 0$ for $\rho (x)$ and $x_*$. The resulting saddle point configuration is
\bea \label{eqn:saddleFsq}
 \rho (x)  \ = \  
{2x \over x_\star^2}
 , \qquad 
x_\star^{2} \ = \ \frac{\omega_{\mathrm{tot}}^{2}}{2 (8 - N_{\rm f})} \, .
\eea
Evaluating \eqref{Eqn:Fasymsq} at this saddle gives the well-known result for the free energy~\cite{Alday:2014rxa,Alday:2014bta}
\ie\label{eqn:FsqlargeN}
 F \ &= \  - \frac{\sqrt{2} \, \pi \omega_{\mathrm{tot}}^{3}}{15 \, \omega_1\omega_2\omega_3 \sqrt{8-N_{\rm f}}} {N^{5/2}} + {\cal O}( N^{3/2} ) \,.
\fe
We determine the round-sphere free energy to be
\ie
\label{F0LargeNft}
F_0 \ &= \  - \frac{9 \sqrt{2} \, \pi}{5 \sqrt{8-N_{\rm f}}} {N^{5/2}} + {\cal O}( N^{3/2} ) \,,
\fe
and extract the conformal central charge from the $\omega_i$-dependence of the free energy by the relation~\eqref{CTRelation} to be
\ie\label{CTlargeNft}
C_T \ &= \  \frac{1152 \sqrt{2}}{\pi{\sqrt{8-N_{\rm f}}}} N^{5/2} + {\cal O}( N^{3/2} ) \,.
\fe

\subsection{Flavor central charges}

\subsubsection{Fundamental hypermultiplet symmetry}
\label{Sec:CJSOflav}

To compute the flavor central charge for ${\rm SO}(2N_{\rm f})$, let us now deform the round-sphere partition function (with squashing parameters $\omega_i = 1$), by giving masses to the $N_{\rm f}$ fundamental hypermultiplets. For simplicity, we choose the following mass matrix (see \eqref{eqn:hyperMassAction} and the discussion in Section~\ref{Sec:En}),
\ie
\label{MassMatrixSO}
 \cM \ = \ m_\mathrm{f} \,\left( \ii \sigma_2 \otimes \mathbbm{1}_{N_{\rm f}} \right)& \in &  \mathfrak{so}\left( 2N_{\rm f} \right) \,,
\fe
where $\mathbbm{1}_{N_{\rm f}}$ is the $N_{\rm f} \times N_{\rm f}$ identity matrix, and $\sigma_2$ is the second Pauli matrix. 
By adding this particular mass term to the perturbative partition function, the localized action \eqref{freeenergy} is modified by the replacement
\ie
G_{H}(\pm\lambda_i \mid \vec{\omega})
\ \to \
G_{H}(\pm\lambda_i + m_{\rm f} \mid \vec{\omega}) \, .
\fe
Since the addition of this mass term only changes the saddle point to subleading order in $1/N$, we may simply evaluate the mass term on the saddle point configuration \eqref{eqn:saddleFsq} to find the modification to the free energy,
\ie
\Delta F \ &= \  - { \pi m_\mathrm{f}^{2} } N_{\rm f} N^{3/2} \int_0^{x_\star} \diff x \, \rho(x) x+ {\cal O}( N^{1/2} )  
\\
\ &= \ - \frac{\sqrt{2}\pi  m_\mathrm{f}^2  N_{\rm f}}{\sqrt{8-N_{\rm f}}} N^{3/2}+ {\cal O}( N^{1/2} ) \,.
\fe
The leading corrections of order $N^{1/2}$ come from both the continuum approximation and further terms in the series expansion in $m_{\rm f}$.
Using
\ie
{I}_{{\bf vec}({\rm SO}(2N_{\rm f}))} \ = \ 1, \qquad \tr_{{\bf vec}({\rm SO}(2N_{\rm f}))}({\cal M}^2) 
\ = \ 2 N_{\rm f} m_{\rm f}^2 \, ,
\fe
in the relation~\eqref{CJRelation2}, we determine the flavor central charge for ${\rm SO}(2N_{\rm f})$ to be
\ie\label{CJSOflav}
C_{J}^{{\rm SO}(2N_{\rm f})} 
 \ = \  \frac{256\sqrt{2}}{3 \pi {\sqrt{8-N_{\rm f}}}} N^{3/2}  + {\cal O}( N^{1/2} )\,.
\fe
In this case, the embedding index into the enhanced flavor symmetry group at the ultraviolet fixed point is simply one, and therefore,
\ie\label{SJSOeqEN}
C_{J}^{E_{N_{\rm f}+1}} \ = \ C_{J}^{{\rm SO}(2N_{\rm f})} \,.
\fe

\subsubsection{Instantonic symmetry}
\label{Sec:FlavorU1J}

Next, we compute the flavor central charge for the instantonic ${\rm U}(1)_{\rm I}$ symmetry. The flavor symmetry ${\rm SO}(2N_{\rm f}) \times {\rm U}(1)_{\rm I}$ in the infrared is expected to enhance to $E_{N_{\rm f}+1}$ at the ultraviolet superconformal fixed point. This implies that the value of $C_J^{E_{N_{\rm f}+1}}$ predicted by the $C_J$ of the ${\rm U}(1)_{\rm I}$ subgroup and that of the ${\rm SO}(2N_{\rm f})$ subgroup should agree (with the embedding indices properly accounted for).  For general $N$, this statement is difficult to prove, as it requires re-summing all instanton contributions.\footnote{In~\cite{Chang:2017cdx}, the present authors obtained some numerical evidence for the agreement in the case of $N = 1$.
}
However, we expect to obtain a precise agreement in the large $N$ limit, since such non-perturbative contributions are exponentially suppressed as discussed before.

To compute the $C_J$ for ${\rm U}(1)_{\rm I}$, we simply keep track of the dependence of the perturbative partition function on the instanton particle mass $m_{\rm I}$, which is related to the Yang-Mills coupling by
\ie
m_{\rm I} \ = \ \frac{4 \pi^{2}}{g_{\mathrm{YM}}^{2}} \,,
\fe
and only appears in the classical piece. The large $N$ localized action including the classical piece reads
\ie
\label{LargeNExponent}
F[\rho] \ &= \
- N^{5/2} \int_{0}^{x_\star} \diff x \, \rho(x) \int_{0}^{x_\star} \diff y \, \rho(y)  \left[ \frac{9\pi}{8} \left( x+y+|x-y| \right)  - \frac{(8-N_{\rm f}) \pi}{3 \, \omega_1\omega_2\omega_3} x^{3}\right]
\\
& \qquad
+2\pi m_{\rm I} N^{2}\int_0^\infty \diff x\, \rho(x)x^2
-\frac{\pi (16+N_{\rm f})}{4} N^{3/2}\int_0^\infty \diff x\,\rho(x)x \,,
\fe
up to corrections that are non-perturbative in $1/N$, since the asymptotic expansions \eqref{expansionhyper} of $\log S_{3}$ have no subleading power law correction.

If we naively proceed as in the previous section, by evaluating the additional instanton mass term on the saddle point \eqref{eqn:saddleFsq} without mass deformations, then we immediately run into a problem, which is that the dependence on $m_{\rm I}$ truncates at linear order. Thus, to find the $C_J$ for ${\rm U}(1)_{\rm I}$, we must include corrections to the saddle that are subleading in $1/N$. We assume that $\rho(x)$ has support on the interval $\left[ x_1,x_2 \right]$, with $x_{i} \geq 0$. The saddle point equation with inclusion of the instanton mass $m_{\rm I}$ reads
\ie\label{eqn:SPE}
0 \ &= \ -{9\pi\over 4}N^{5/2}\int_{x_1}^{x_2} \diff y\,\rho(y)(|x-y|+x+y )
\\
& \qquad + {\pi(8-N_{\rm f})\over 3}N^{5/ 2}x^3 + 2\pi m_{\rm I} N^{2} x^2 -{\pi (16+N_{\rm f}) \over 4}N^{3/2}x+\mu \,,
\fe
where as before $\mu$ is the Lagrange multiplier for the normalization of $\rho(x)$. Taking the derivative twice with respect to $x$ shows that $\rho(x)$ is a linear function. The saddle point equation can be straightforwardly solved to give
\bea
\rho(x) \ = \ \frac{2x}{a} +b\,, \quad x \in [x_1, x_2]\,,
\eea
where\footnote{To leading order at large $N$, we recover the previous saddle point \eqref{eqn:saddleFsq}.
}
\ie
 a \ &= \  {9(8-N_{\rm f})\over 4},\qquad b \ = \ {8m_{\rm I}\over 9 \sqrt{N}} \,,\\
x_1 \ &= \  {\sqrt{16m_{\rm I}^2+ (8-N_{\rm f})(16+N_{\rm f})}-4 m_{\rm I}\over 2(8-N_{\rm f})\sqrt{N} }\,,\\
x_2 \ &= \  {\sqrt{16m_{\rm I}^2+(16+N_{\rm f}+18N)(8-N_{\rm f})}-4m_{\rm I}\over  2(8-N_{\rm f})\sqrt{N}}\,.
\fe
We can now compute the leading order large $N$ free energy $F$ at the saddle point,
\ie
F \ &= \ -{9\sqrt{2}\pi N^{5/2}\over 5\sqrt{8-N_{\rm f}}}+{9\pi m_{\rm I} N^2 \over 2(8-N_{\rm f})} - {4\sqrt{2} \pi m_{\rm I}^2 N^{3/2} \over (8-N_{\rm f})^{3/2}}
+{16\pi m_{\rm I}^3 N \over 3(8-N_{\rm f})^2} + {\cal O}(N^{1/2}) \,.
\label{FmIexp}
\fe
The leading correction of order $N^{1/2}$ comes from the continuum approximation of the discrete Coulomb branch parameters $\lambda_i$ by $\rho(x)$.  By regarding the discrete sum over $i = 1, 2, \dotsc, N$ as a (midpoint) Riemann sum for the integral over $x$, we estimate the error of the continuum approximation to be of order $N^{-2}$ relative to the leading $N^{5/2}$.  In comparison, the correction coming from the Gaussian integral around the saddle point is of order $\log N$.

There appears to be an immediate problem.  The term of order $m_{\rm I}$ would imply a non-vanishing sphere one-point function of the flavor currents, which violates conformal invariance. However, as discovered in~\cite{Chang:2017cdx} and reviewed in Section~\ref{Sec:counter-terms}, there are ambiguities in the free energy due to counter-terms. In particular, upon localization on the five-sphere, the mixed Chern-Simons counter-terms contribute \emph{real} pieces at order $m_{\rm I}$ and $m_{\rm I}^{3}$~\cite{Chang:2017cdx}. Thus, we may pick a regularization scheme to remove the terms with linear and cubic dependence on $m_{\rm I}$ in~\eqref{FmIexp}, to properly preserve conformal symmetry at the ultraviolet fixed point.

We can now compute the flavor central charge for the instantonic ${\rm U}(1)_{\rm I}$ current by the relation \eqref{CJRelation1},
\ie\label{eqn:U1CJ}
C_J^{{\rm U}(1)_{\rm I}} \ = \ {1024\sqrt{2} \over 3\pi(8-N_{\rm f})^{3/2}} N^{3/2} + {\cal O}(N^{1/2})\,.
\fe
The flavor central charge of the enhanced $E_{N_{\rm f}+1}$ flavor symmetry is related to that of the instantonic ${\rm U}(1)_{\rm I}$ by the embedding index \eqref{Embedding}, and we conclude that
\ie\label{CJEfromU1I}
C_J^{E_{N_{\rm f}+1}} \ = \ {8-N_{\rm f}\over 4}C_J^{\rm {\rm U}(1)_{\rm I}} \ = \ {256\sqrt{2} \over 3\pi \sqrt{8-N_{\rm f}} } N^{3/2} + {\cal O}(N^{1/2})\,.
\fe
This precisely agrees with the flavor central charge for ${E_{N_{\rm f}+1}}$ computed using the ${\rm SO}(2N_{\rm f})$ mass deformation given in~\eqref{CJSOflav} and~\eqref{SJSOeqEN}, and confirms the flavor symmetry enhancement in Seiberg theories.

\subsubsection{Mesonic symmetry
 }

Finally, we compute the flavor central charge for the mesonic ${\rm SU}(2)_{\rm M}$ symmetry. We deform the five-sphere partition function by a mass for the antisymmetric hypermultiplet, which transforms in the fundamental of ${\rm SU}(2)_{\rm M}$. As before, we couple it to a background vector multiplet with the following choice of mass matrix,
\ie
\label{MassMatrixSU}
\cM \ = \ \ii m_{\mathrm{as}} \, \sigma_{3} \ \in \ \mathfrak{su}(2)_{\rm M} \,.
\fe
Then the localized action \eqref{freeenergy} is modified by the replacement
\ie
G_{H} ( \lambda_i \pm \lambda_j \mid \vec{\omega}) \ \to \ {1\over 2}G_{H} ( \lambda_i \pm \lambda_j \pm m_{\rm as} \mid \vec{\omega}) \,.
\fe
Again, we shall use the asymptotics for the triple sine functions as given in~\eqref{expansionhyper}. The leading order modification to the large $N$ localized action \eqref{Eqn:Fasymsq}, due to the addition of the mass for the antisymmetric hypermultiplet, then reads
\ie
\Delta F[\rho] \ &= \  - \frac{\pi }{2 } N^{5/2} m_{\mathrm{as}}^{2}\int_{0}^{x_{\star}} \diff x \, \rho(x) \int_{0}^{x_{\star}} \diff y \,\rho(y) \, \bigg\{ (x+y) 
\\
& \hspace{1in} + \left[ (x-y) \theta (x>y)
+ (y-x) \theta (y>x) \right]
\bigg\} + {\cal O}( N^{3/2} )\,,
\fe
where $\theta$ denotes the Heaviside theta function, and the leading corrections of order $N^{3/2}$ come from further terms in the series expansion in $m_{\rm as}$. This modifies the saddle point configuration to
\ie
\rho(x) \ &= \ {2x \over x_\star^2} \,, \qquad 
x_\star^2 \ = \ \frac{9 + 4m_{\rm as}^2}{2 \left( 8-N_{\rm f} \right)} \,.
\fe
Consequently, the large $N$ free energy is given by
\ie
F \ &= \ - \frac{\sqrt{2} \pi   \left(9 + 4 m_{\mathrm{as}}^{2}\right) ^{3/2}}{15 {\sqrt{8-N_{\rm f}}} } N^{5/2} + {\cal O}(N^{3/2}) \,.
\fe
As before, we apply the relation \eqref{CJRelation2} with
\ie
{I}_{{\bf fund}({{\rm SU}(2)})} \ = \ {1 \over 2}, \qquad \tr_{{\bf fund}({{\rm SU}(2)})} ({\cal M}^2) \ = \ 2 m_{\rm as}^2,
\fe
to compute the flavor central charge of the mesonic ${\rm SU}(2)_{\rm M}$ flavor symmetry,
\ie
C_{J}^{{\rm SU}(2)_{\rm M}}
\ &= \ \frac{256 \sqrt{2}}{5 \pi {\sqrt{8-N_{\rm f}}} } N^{5/2}+ {\cal O}( N^{3/2} )  \,. \label{CJSU2m}
\fe
This large $N$ formula for the mesonic flavor central charge is in fact equal to that for the R-symmetry flavor central charge. The latter is related to the conformal central charge by superconformal Ward identities, since the R-symmetry currents reside in the same superconformal multiplet as the stress tensor. The agreement of these two flavor central charges at large $N$ is expected from the dual supergravity perspective, where ${\rm SU}(2)_{\rm M}$ and ${\rm SU}(2)_{\rm R}$ combine to form the ${\rm SO}(4)$ isometry of the internal four-hemisphere (see Section~\ref{Sec:Holography}), and are exchanged under a frame rotation as described in Section~\ref{Sec:MesonicSymmetry}.  However, this agreement is obscure from field theoretic considerations, and fails at finite $N$.

\subsection{Finite $N$ numerics}

We have argued that the perturbative partition function is {\it exact} in the large $N$ limit.  Away from large $N$, instanton corrections are no longer suppressed, and computing the exact partition function or central charges becomes a difficult task. However, in~\cite{Chang:2017cdx}, the present authors observed that for Seiberg theories of rank one -- which is the opposite of large $N$ -- the instanton contributions to the central charges as extracted from the supersymmetric five-sphere partition function are also small. Combining these two facts, it is natural to expect that instanton contributions are in fact small for all $N$.

\begin{figure}[H]
\centering
\subfloat{
\includegraphics[width=.45\textwidth]{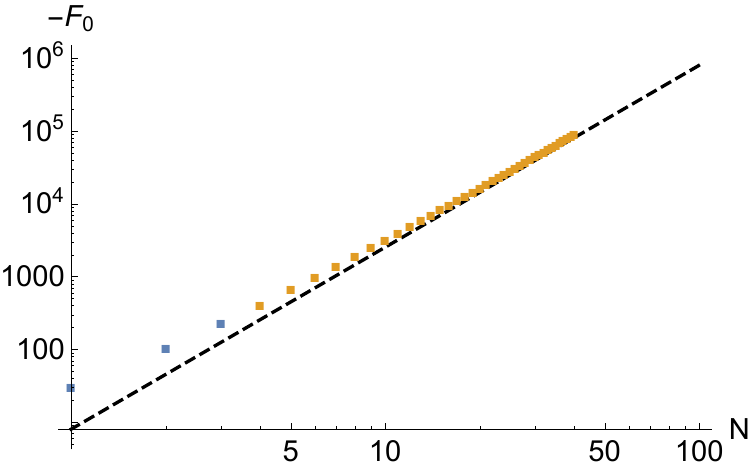}
}
\quad
\subfloat{
\includegraphics[width=.45\textwidth]{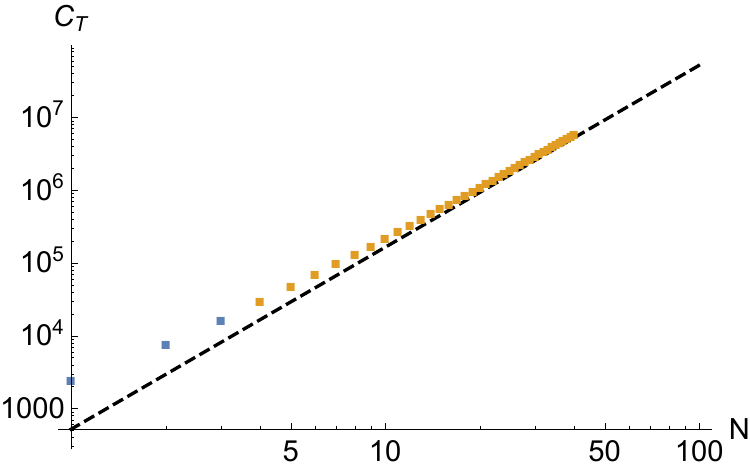}
}
\\
\subfloat{
\includegraphics[width=.45\textwidth]{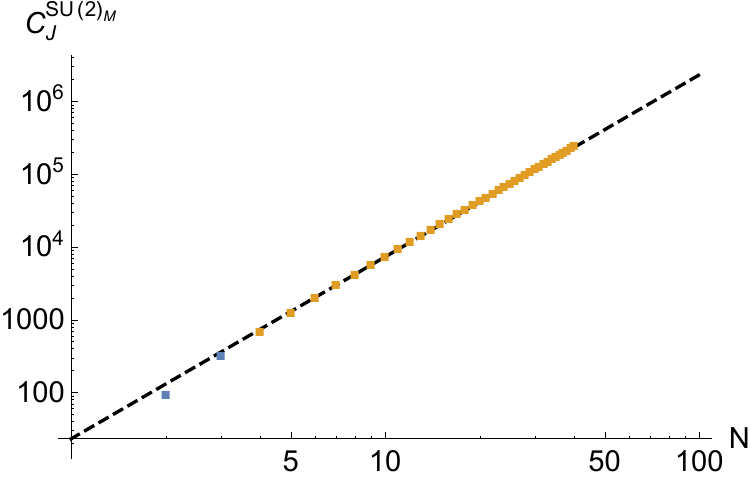}
}
\quad
\subfloat{
\includegraphics[width=.45\textwidth]{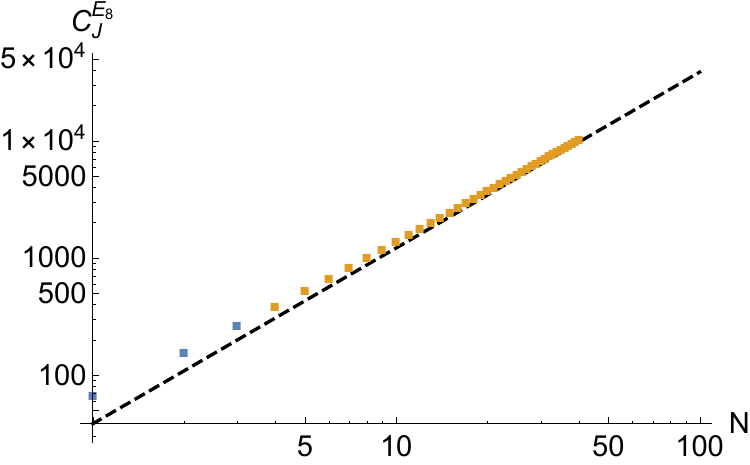}
}
\caption{Juxtapositions for various physical quantities in the Seiberg $E_8$ theory of results obtained by numerically computing the perturbative partition function (squares), and by the large $N$ formula (dashed line). The physical quantities considered here are the round sphere free energy $-F_0$, the conformal central charge $C_T$, the mesonic flavor central charge $C_J^{{\rm SU}(2)_{\rm M}}$, and the exceptional flavor central charge $C_J^{E_8}$. The numerical computations are done by direct integration up to $N = 3$, and by saddle point approximation up to $N = 40$.}
\label{Fig:Juxta}
\end{figure}

Under this assumption, we can approximate the exact partition function by the perturbative formula, and numerically compute the round-sphere free energy and the central charges at finite values of $N$. For $1 \leq N \leq 3$, the integral \eqref{eqn:USpPartition} can be computed straightforwardly by direct numerical evaluation; for $1 \leq N \leq 40$, we can estimate the integral by a saddle point approximation. In the absence of a large parameter, the latter is {\it a priori} illegal.  However, after explicitly computing the round sphere free energy and the various central charges, we find that the saddle point approximation agrees with direct numerical integration even for $N = 3$ to within 1\% (these numerical results are presented in detail in Appendix~\ref{App:Numerics}).  Thus, we believe that this approximation can in fact be trusted for $N \geq 3$.  These results suggest that in situations where only approximate values of these quantities are needed, such as for numerical bootstrap, the finite $N$ saddle point approximation serves as an efficient method to perform the perturbative localization integral \eqref{eqn:USpPartition}.  In Figure~\ref{Fig:Juxta}, we juxtapose the results of direct numerical integration, finite $N$ saddle point approximation, and the large $N$ formula, in the case of the Seiberg $E_8$ theory.

\section{Central charges from supergravity}
\label{Sec:Holography}

We now move towards a study of the central charges in the holographic duals of the Seiberg theories.  As reviewed in Section~\ref{Sec:En}, the Seiberg theories can be constructed by a system of D4-branes probing an orientifold singularity in type I' string theory\cite{Seiberg:1996bd}, the configuration of which is given in Table~\ref{table:BraneDirections}.  In summary, the configuration has $N_{\rm f} < 8$ D8-branes coinciding with an O8, on top of which lie $N$ D4-branes.  The decoupling limit suggests a holographic correspondence between Seiberg theories and type I' string theory on a background of the form $\cM_{6} \times_{\rm w} {\rm HS}^{4}$, a warped product of a four-hemisphere with a asymptotically locally AdS$_6$ space $\cM_{6}$~\cite{Ferrara:1998gv,Brandhuber:1999np,Bergman:2012kr}.  In an appropriate low energy regime, the gravity side is well approximated by type I' supergravity~\cite{Polchinski:1995df}; in particular, the region between D8-branes is described by Romans massive type IIA supergravity~\cite{Polchinski:1995df,Romans:1985tz}.  This latter regime is what we mainly consider in the following.

This particular holographic correspondence has thus far been tested in various ways.  Initial checks include comparisons of the symmetries on both sides~\cite{Ferrara:1998gv, Brandhuber:1999np}, and more recent checks include comparisons of the entanglement entropy~\cite{Jafferis:2012iv} and the free energy/Wilson lines in the case of ${\cal M}_6 = \text{AdS}_{6}$~\cite{Assel:2012nf}, 
as well as for more general asymptotically locally $ \text{AdS}_{6}$ spacetimes~\cite{Alday:2014rxa,Alday:2014bta,Alday:2015jsa}.

\subsection{Central charges and couplings}

The conformal central charge $C_T$ and the flavor central charges $C_J$ in a $d$-dimensional superconformal field theory with a weakly coupled AdS$_{d+1}$ dual are related to specific coupling constants in the supergravity action. The leading order derivative expansion of the weakly coupled bulk effective action is schematically given by
\ie
\label{eqn:AdS6action}
S_{d+1} & = \int \diff^{d+1} x \sqrt{-g_{d+1}} \left[ \frac{1}{2\kappa_{d+1}^{2}} \left( R_{d+1} - 2 \Lambda_{d+1} \right) - \frac{1}{2 e^{2}} {\rm Tr} \left( F^{(d+1)}_{\mu\nu}F^{(d+1)}{\,}^{\mu\nu} \right)  + \cdots \right] \,,
\fe
where $R_{d+1}$ is the $(d+1)$-dimensional Ricci scalar, $F^{(d+1)}_{\mu\nu}$ is the $(d+1)$-dimensional field strength for the gauge field which is sourced by the flavor symmetry current at the conformal boundary, $\Tr(\cdot)$ is the Killing form defined in \eqref{Killing}, and finally,
\ie
\Lambda_{d+1} \ = \ -\frac{d (d-1)}{2 \ell^{2}}
\fe
is the $(d+1)$-dimensional cosmological constant, with $\ell$ the radius of AdS$_{d+1}$. The gravitational coupling $\kappa_{d+1}$ and the gauge coupling $e$ are related to the central charge $C_T$ and the flavor central charge $C_J$ by~\cite{Freedman:1998tz,Penedones:2016voo}
\bea
C_{T}^{(d)} & \ = \ & \frac{4 \pi^{d/2} \Gamma (d+2)}{(d-1) \Gamma \left( \frac{d}{2} \right)^{3}} \frac{\ell^{d-1}}{\kappa_{d+1}^{2}} \label{eqn:CTgravgen}\,,\\
C_{J}^{(d)} & \ = \ & \frac{2^{d+1} (d-2) \pi^{(d-1)/2} \Gamma\left( \frac{d+1}{2} \right)}{\Gamma\left( \frac{d}{2} \right)^{2}} \frac{\ell^{d-3}}{e^{2}} \label{eqn:CJgravgen}\,.
\eea
These relations allow us to extract $C_T$ and $C_J$ from the supergravity duals of the Seiberg theories. The key is then to determine the precise values of the couplings $\kappa_{d+1}$ and $e$ that properly embed the supergravity into the aforementioned type I' string theory background.

\subsection{Massive IIA supergravity solutions}
\label{Sec:MassiveIIASeiberg}

Let us first review the AdS$_6 \times {\rm HS} ^{4}$ solutions in massive IIA supergravity~\cite{Brandhuber:1999np}. In the string frame, the bosonic part of the ten-dimensional massive IIA action reads
\ie
\label{SIIA}
S_{10}^{\rm IIA} \ & = \ - \frac{1}{2 \kappa_{10}^{2}} \int \diff^{10}x \, \sqrt{-g_{10}}\left[ e^{-2\Phi} \left( R_{10}+4 \partial_\mu \Phi \partial^{\mu} \Phi\right) - \frac{1}{2 \cdot 6! } \left| F_{6} \right|^{2} - \frac{1}{2} m_{\text{IIA}}^{2} \right] \,,
\fe
where $g_{10}$ is the ten-dimensional metric, $R_{10}$ the corresponding Ricci scalar, and
\ie
\kappa_{10} \ = \ 8\pi^{7/2}\ell_s^4
\fe
is the ten-dimensional gravitational coupling expressed in terms of the string length $\ell_s$. Furthermore, we denote by $\Phi$ the ten-dimensional dilaton, $F_6$ the six-form flux, and $m_{\rm IIA}$ the Romans mass. The string frame metric for the AdS$_{6} \times {\rm HS}^{4}$ background is explicitly given by
\ie\label{eqn:10metric}
 \diff s_{10}^{2} \ &= \  \frac{1}{\sin^{1/3}\A } \left[ L^{2} \diff s^{2}_{\widehat{\text{AdS}_{6}}}+R^{2} \left( \diff \alpha^{2} + \cos^{2} \alpha \ \diff s^{2}_{ {\rm S}^{3}} \right)\right] \,,
\fe
where $L$ is the AdS$_6$ radius, $R$ is the ${\rm HS}^4$ radius, and $\alpha \in (0, \pi /2 ]$. We denote by $\diff s^{2}_{\widehat{\text{AdS}_{6}}}$ the standard metric of the unit-radius AdS$_6$, and by $\diff s^{2}_{ {\rm S} ^{3}}$ the unit ${\rm S}^{3}$-slices of the four-hemisphere,
\ie\label{Eqn:S3coords}
\diff s^{2}_{ {\rm S} ^{3}} \ &= \  \frac{1}{4} \left[ \diff \theta_1^{2} + \sin^{2} \theta_1 \, \diff \theta_2^{2}+ \left( \diff \theta_3 - \cos \theta_1 \, \diff \theta_2 \right)^{2}\right] \,,
\fe
with $\theta_1 \in [0, \pi]$, $\theta_2 \in [0,2 \pi)$, and $\theta_3 \in [0, 4\pi)$. Note that the isometry group of ${\rm S}^{3}$ is ${\rm SO}(4) \cong {\rm SU}(2)_{\rm M} \times {\rm SU}(2)_{\rm R}$, where the former ${\rm SU}(2)_{\rm M}$ factor corresponds in the dual five-dimensional superconformal field theory to the mesonic global symmetry, and the latter to the R-symmetry. 

Next, we relate this general ansatz to the parameters of the string theory setup. Each D4-brane has tension
\ie\label{eqn:T4}
T_4 \ = \ 
(2\pi)^{-4} \ell_s^{-5} \, ,
\fe
and therefore, $N$ D4-branes source the four-form flux (Hodge dual to $F_6$ in the above action),
\ie\label{4Form}
F_4 \ &= \  
\frac{10}{9} \pi N \ell_{s}^{3}
 \cos^{3} \alpha \sin^{1/3}\alpha \sin \theta_{1} \diff \alpha \wedge \diff \theta_1\wedge \diff\theta_2 \wedge \diff \theta_3\,,
\fe
such that the total D4-brane charge is
\ie\label{Eqn:D4branechargeSeiberg}
\frac{1}{2\kappa_{10}^{2}} \int_{ \rm HS ^{4}} F_4 \ = \ T_4 N \,.
\fe
Similarly, each D8-brane has charge
\ie
\mu_8 \ = \ (2\pi)^{-9/2}\ell_{s}^{-5} \, ,
\fe
and so $N_{\rm f}$ D8-branes and one O8-brane source the Romans mass,
\ie
 m_{\rm IIA} \ &= \  \sqrt{2} (8-N_{\rm f}) \mu_8 \kappa_{10} \ = \ \frac{8-N_{\rm f}}{2\pi \ell_s} \,.
\fe

With the above input from string theory, the supergravity Killing spinor equations fix the AdS$_6$ radius and the HS$^4$ radius to be\footnote{We use $L$ to denote the AdS$_6$ radius in the full ten-dimensional background, and reserve $\ell$ for the AdS$_6$ radius in the reduced six-dimensional background that appears later.
}
\ie\label{L/lsinN}
 \frac{L^{4}}{\ell_{s}^{4}} \ &= \  \frac{18 \pi^{2} N}{(8-N_{\rm f})} \,, \qquad
R \ = \ {2 L \over 3}\,,
\fe
and the dilaton profile to be
\ie\label{eqn:solsIIAPhim}
e^{-2 \Phi} \ &= \  \frac{3}{2 \sqrt{2} \pi }  N^{1/2} (8-N_{\rm f})^{3/2} \sin ^{5/3}\alpha\,.
\fe
The above solution in massive IIA preserves sixteen supersymmetries. 

Next, following~\cite{Brandhuber:1999np}, we identify a regime where type IIA string theory is effectively described by supergravity.  The dilaton diverges at the boundary $\alpha = 0$, which is related to the fact the Yang-Mills coupling in the five-dimensional gauge theory diverges as we approach the ultraviolet superconformal fixed point.  The curvature also diverges as $\alpha \to 0$,\footnote{The $R_{10}$ here is the string frame Ricci scalar.  The Einstein frame Ricci scalar also diverges at $\A = 0$. In the dual heterotic picture~\cite{Polchinski:1995df}, while the dilaton is small everywhere, the curvature (in both string and Einstein frames) is large near $\A = 0$.
}
\ie
R_{10} \ell^{2}_{s} \ &\propto \ (8-N_{\rm f})^{1/2} N^{-1/2} \sin^{-5/3} \alpha \,.
\fe
Nonetheless, when $N$ is large, there is a regime 
\ie
\sin\alpha \ \gg \ N^{-3/10}
\fe
where both the curvature and the dilaton are small, and supergravity can be trusted.  Note that the D0-branes have masses of order
\ie
 M_{\rm D0} \ \sim \ N^{1/2} \sin^{2/3}\alpha \,,
\fe
and therefore do not renormalize the supergravity action in this regime.  This is the bulk counterpart of the argument presented at the beginning of Section~\ref{Sec:LargeN} for the suppression of instanton contributions.

\subsection{Conformal central charge from supergravity}
\label{Sec:CTfromGravitySeiberg}

We start by computing the conformal central charge from supergravity. In order to derive the effective six-dimensional gravitational coupling constant, we perform a Kaluza-Klein reduction on the ten-dimensional massive IIA action. A consistent truncation of massive IIA supergravity to Romans $F(4)$ gauged supergravity~\cite{Romans:1985tw} was found in~\cite{Cvetic:1999un}, and the holographic correspondence for the six-dimensional Euclidean effective theory was discussed in some detail in~\cite{Alday:2014rxa, Alday:2015lta}. To make precise contact with the latter references, we denote by $\diff s^{2}_{\text{AdS}_{6}}$ the AdS$_6$ metric of radius 
\ie
\ell \ = \ {3 \over \sqrt{2}} \, ,
\fe
and rewrite the above ten-dimensional metric~\eqref{eqn:10metric} as
\ie\label{eqn:10metricModified}
 \diff s_{10}^{2} \ &= \  \frac{L^{2}}{\ell^2\sin^{1/3}\A } \left[  \diff s^{2}_{\text{AdS}_{6}}+ {4 \over 9} \ell^2 \left( \diff \alpha^{2} + \cos^{2} \alpha \ \diff s^{2}_{{\rm S}^{3}} \right)\right] \,.
\fe

In the background discussed above, the massive IIA action $S_{10}^{\rm IIA}$ in \eqref{SIIA} truncates to an action for Romans $F(4)$ supergravity,
\ie
\label{eqn:IIAaction}
S_6^{F(4)} \ = \ -\frac{N^{5/2}}{30 \pi^2 \sqrt{2} \sqrt{8-N_{\rm f}}} \int_{\text{AdS}_6} \diff^{6}x \sqrt{-g_{6}} \, \left( R_{6} - 2 \Lambda_{6} \right) \,,
\fe
where we have used the solution~\eqref{eqn:solsIIAPhim} for the dilaton $\Phi$, the ratio \eqref{L/lsinN}, and denoted by $\Lambda_{6}$ the six-dimensional cosmological constant.\footnote{Note that the ten-dimensional Ricci scalar can be expressed in terms of the six-dimensional one as
\ie
R_{10} = \frac{\ell^2\sin^{1/3} \alpha }{L^2} R_{6} + \cdots\,,
\fe
where the ellipses denote additional terms due to the warping factor.
}
Comparing the reduced six-dimensional action \eqref{eqn:IIAaction} with the canonical form \eqref{eqn:AdS6action}, we read off the value of the six-dimensional gravitational coupling constant, 
\ie
\kappa_{6}^{2} \ &= \ \frac{15 \pi^{2}  \sqrt{2(8-N_{\rm f})}}{ N^{5/2}}  \,.
\fe
The conformal central charge in the dual superconformal field theory is given by~\eqref{eqn:CTgravgen} to be
\ie \label{eqn:CTsugra}
 C_T \ &= \  \frac{1152 \sqrt{2}}{\pi} \, \frac{N^{5/2}}{\sqrt{8-N_{\rm f}}} \,.
\fe
The free energy of the effective six-dimensional supergravity for AdS$_{6}$ can also be evaluated\footnote{We refer to~\cite{Alday:2014rxa, Alday:2015lta} for more details on the counter-terms, which we just denoted collectively as $I_{\mathrm{ct}}$ here. Note that in the case of even-dimensional bulk (odd-dimensional boundary), the finite part of the free energy is scheme-independent due to the absence of logarithmic divergence.
}
\ie \label{eqn:6daction}
F_0 \ &= \ - \frac{1}{2\kappa_6^2} \int_{\text{AdS}_6} \diff^{6}x \sqrt{-g_{6}}\, \left( R_{6} - 2 \Lambda_{6} \right) + I_{\mathrm{ct}}
\\
\ &= \ -\frac{54 \pi ^3}{\kappa_6^2} \ = \ - \frac{9 \sqrt2 \pi}{5} \, \frac{N^{5/2}}{\sqrt{8-N_{\rm f}}} \,,
\fe
where we used $\Lambda_{6} = \tfrac{1}{3}R_{6}$.\footnote{The cosmological constant $\Lambda_{d}$ for $d$-dimensional AdS space is related to the Ricci scalar $R_d$ by
\ie
\Lambda_{d} \ = \ \frac{d-2}{2d} R_{d} \,.
\fe
}
These precisely agree with the round sphere free energy~\eqref{F0LargeNft} and the conformal central charge~\eqref{CTlargeNft} computed by localization in the large $N$ field theory.

\subsection{Flavor central charges from supergravity}
\label{Sec:FlavCJSeiberg}

We now turn our attention to extracting from supergravity the flavor central charges of the Seiberg theories.

\subsubsection{Mesonic symmetry}
\label{Sec:MesonicSymmetry}

Let us first consider the mesonic ${\rm SU}(2)_{\rm M}$ symmetry. The corresponding gauge field in the dual six-dimensional effective supergravity is on equal footing with the ${\rm SU}(2)_{\rm R}$ R-symmetry gauge field. We presently explain this in more details.

The internal $\rm S^3\cong SU(2)$ in $\rm HS^4$ has $\rm SU(2)\times SU(2)$ isometry generated by left and right group multiplications. Correspondingly, there are left- and right-invariant one-forms $\sigma^i$ and $\widetilde\sigma^i$. They define two vielbeins on $\rm S^3$,
\ie
g_{mn}({\rm S^3}) \ = \ \sigma^i_m \sigma^i_n \ = \ \widetilde\sigma^i_m \widetilde\sigma^i_n,
\fe
which are related by a local $\rm SO(3)$ frame rotation. To incorporate fluctuations of the six-dimensional $SU(2)_R$ gauge fields, the reduction ansatz for the ten-dimensional metric involves shifting $\sigma^i$ by six-dimensional gauge connections~\cite{Cvetic:1999un}
\ie
\sigma^i \ \to \ \sigma^i - A_R^i.
\fe
Consistency with the ten-dimensional equations of motions requires the R-R four-form flux $\hat F_{(4)}$ to be adjusted accordingly and one needs to keep track of this carefully to obtain the correct six-dimensional gauge field kinetic term. To turn on $\rm SU(2)_M$ gauge field fluctuations, we should instead shift the right invariant $\rm SU(2)$ one-forms 
\ie
\widetilde\sigma^i \ \to \ \widetilde\sigma^i - A_M^i.
\fe
Since this is simply related to the former analysis for $\rm SU(2)_R$ by a local frame rotation, we are guaranteed to find the same reduction ansatz at linearized order, and thus the same gauge field kinetic term after Kaluza-Klein reduction. It follows that $C_J^{{\rm SU}(2)_{\rm M}}$ and $C_J^{{\rm SU}(2)_{\rm R}}$ are equal in the large $N$ limit.

 In the Euclidean action for Romans $F(4)$ gauged supergravity~\cite{Alday:2014rxa, Alday:2015lta},
\ie
S_{6}^{F(4)} \ &= \  - \frac{1}{2\kappa_6^2} \int \left[ R_6 * 1 -\frac{1}{2} X^{-2} F^{i} \wedge * F^{i} + \cdots \right] \,,
\fe
the isometries of the internal four-hemisphere manifest as the gauge fields $A^i$ that give rise to the field strengths
\ie
F^{i} \ = \ \diff A^i -{1\over 2}\epsilon^{ijk}A^j\wedge A^k \,,
\fe
where $i,j,k=1,2,3$ are the adjoint indices for the ${\rm SU}(2)_{\rm R}$.  The gauge coupling constant for either the mesonic symmetry or the R-symmetry are simply related to the six-dimensional gravitational coupling constant,
\ie
e^{2}_{\rm M} \ = \ 2\kappa_6^2 \ = \ \frac{30 \pi^{2} \sqrt{2(8-N_{\rm f})}}{ N^{5/2}} \,.
\fe
The mesonic flavor central charge is determined by the formula~\eqref{eqn:CJgravgen} to be 
\ie \label{eqn:CJmessugra}
C_{J}^{{\rm SU}(2)_{\rm M}} \ &= \  \frac{256 \sqrt{2}}{5\pi} \frac{N^{5/2}}{ \sqrt{8-N_{\rm f}}}\,,
\fe
which is again in precise agreement with the large $N$ field theory result given in~\eqref{CJSU2m}.

\subsubsection{Fundamental hypermultiplet symmetry}

Next, we compute from supergravity the flavor central charge for the ${\rm SO}(2N_{\rm f}) \subset E_{N_{\rm f}+1}$ subgroup. In the full ten-dimensional string theory, the ${\rm SO}(2N_{\rm f})$ flavor symmetry manifest as gauge fields $G$ supported on the nine-dimensional worldvolume of the $N_{\rm f}$ D8-branes located at $x^{9}=0$ (or $\A = 0$). Thus, let us start with the ten-dimensional type I' action (\Ie,~massive IIA coupled to nine-dimensional gauge fields on the D8-branes)~\cite{Polchinski:1995mt,Polchinski:1995df} (see also~\cite{Bergman:1997py, Matalliotakis:1997qe}),
\ie
S^{{\rm I'}}_{10} \ &= \  - \frac{1}{2 \kappa_{10}^{2}} \int \diff^{10}x \, \sqrt{-g_{10}}\left[ e^{-2\Phi} \left( R+4 \partial_\mu \Phi \partial^{\mu} \Phi\right) - \frac{1}{2 \cdot 6!}  \left| F_{6} \right|^{2} -  \frac{1}{2} m_{\rm IIA}^{2} \right]
\\
& \hspace{.5in} - \frac{\mu_{8}}{\sqrt{2} \kappa_{10}} \sum_{i} \int_{x^{9} \equiv x_i^{9}} \diff^9 x \sqrt{-g_9}\left[ e^{-\Phi} \left( \pi \ell_s^{2} \right)^{2} {\rm Tr}\, g_9^{\m\rho}g_9^{\n\sigma}{G}_{\mu\nu}{G}_{\rho\sigma}+ \cdots\right] \,,
\fe
where the Killing form $\Tr(\cdot)$ is defined in~\eqref{Killing}, $x_{i}^{9}$ are the positions of the D8-branes ($x_{i}^{9}=0$ for $i = 1, \dotsc, N_{\rm f}$, and $x_{i}^{9}=\pi$ for $i = N_{\rm f}+1, \dotsc, 16$), and $\mu_8$ is the coupling constant between the nine-form gauge field and the D8-branes. Finally,
\ie
g_9 \ = \ g_{10} \big|_{x^{9}=x^{9}_{i}}
\fe
is the induced nine-dimensional metric on the D8-branes. Thus, the induced effective six-dimensional action is
\ie
\label{InducedGauge}
& \, \hspace{-.2in} S_{6}^{\text{gauge}}
\\
& \hspace{-.2in} = \, - \frac{\left( \pi \ell_{s}^{2} \right)^{2} \mu_{8}}{\sqrt{2}\kappa_{10}}
\int_{ {\rm S}^{3}}\sqrt{g_{3}}\left[ \left(\frac{L^{2}}{ \sin^{1/3} \alpha}\right)^{5/2} R^{3/2} e^{-\Phi}\right]
\int \diff^{6}x \sqrt{-g_{6}}\,g_6^{\m\rho}g_6^{\n\sigma} {\rm Tr} \,{G}_{\mu\nu} {G}_{\mu\nu}+\cdots \\
& \hspace{-.2in} = \ - \frac{N^{3/2}}{36 \pi ^2 \sqrt{2(8-N_{\rm f})}}
\int \diff^{6}x \sqrt{-g_{6}}\,  g_6^{\m\rho}g_6^{\n\sigma}{\rm Tr}\, {G}_{\mu\nu} {G}_{\rho\sigma}+\cdots \,,
\fe
where we have used the expressions given in Section~\ref{Sec:MassiveIIASeiberg}.
The six-dimensional effective coupling constant can be read off the above equation by comparing with the canonical form~\eqref{eqn:AdS6action},
\ie
\frac{1}{e^{2}} \ &= \   \frac{N^{3/2}}{18 \pi ^2 \sqrt{2(8-N_{\rm f})}}\,,
\fe
which by~\eqref{eqn:CJgravgen} determines
\ie\label{eqn:CJmfsugra}
C_J^{{\rm SO}(2N_{\rm f})} \ &= \  \frac{256 \sqrt{2}}{3 \pi}\frac{ N^{3/2}}{  \sqrt{8-N_{\rm f}}} \,.
\fe
This again precisely agrees with the field theory result~\eqref{CJSOflav}.
 
It is rather surprising that with supergravity we can access information about this last flavor central charge. The ${\rm SO}(2N_{\rm f})$ gauge fields are inherently supported on the (singular) boundary $\alpha =0$, at which our supergravity approximation supposedly breaks down. Miraculously, the factors of $\alpha$ coming from the metric and from the dilaton precisely cancel in the above computation, and we end up with a finite result.

\subsubsection{Instantonic symmetry}
\label{sect:D0CJ}

Let us briefly remark on the gravity dual to the instantonic ${\rm U}(1)_{\rm I}$ flavor symmetry. From the type I' construction of the Seiberg theories, it is clear that the duals of the operators with instanton charges involve D0-branes. However, on the one hand, due to the Wess-Zumino coupling \cite{Green:1996bh}, $m_{\rm IIA} \int a$, the D0-brane worldvolume gauge field $a$ has a tadpole and needs to be cancelled by attaching fundamental strings, whose other ends lie on the D4/D8 branes at $\A=0$. On the other hand, the bulk gauge field $A_{\rm I}$ dual to the ${\rm U}(1)_{\rm I}$ symmetry current descends not just from the R-R one-form $C_1$, as the equations of motions in massive type IIA supergravity require the gauge field $A_{\rm I}$ to appear also in the reduction ansatz of the NS-NS two-form $B_2$ and the R-R three-form $C_3$~\cite{Cvetic:1999un}. As a result, one needs to take into account all the ten-dimensional NS-NS and R-R form fields in order to produce the correct kinetic term for $A_{\rm I}$ in the six-dimensional supergravity.

There is an additional complication due to the nontrivial warping. Unlike the non-warped maximally supersymmetric cases, the (linearized) reduction ansatz for the ten-dimensional form fields now involves tensor harmonics  on $\rm S^4$ dressed by scalar functions of $\alpha$, which are constrained by the ten-dimensional equations of motions.\footnote{We demand that the six-dimensional equations of motion take canonical forms. Then, the reduction ansatz in terms of the six-dimensional fields must solve the ten-dimensional equations of motion provided that the six-dimensional counter-parts are satisfied.
}
Furthermore, due to the divergence of the ten-dimensional dilaton and metric at $\alpha=0$, the integral over $\rm S^4$ may require additional counter-terms at $\alpha=0$ to produce a finite coupling in six dimensions~\cite{Cvetic:1999un}. We plan to come back to these subtle issues in the future.

\section{Orbifold theories and their large $N$ central charges}
\label{Sec:QuiverTh}

We shall now extend our analysis of central charges in five-dimensional superconformal field theories and their gravity duals to a generalization of the Seiberg theories, described in the infrared by quiver gauge theories with ${\rm SU}(N)$ and ${\rm USp}(2N)$ gauge groups. From the type I' string theory perspective, these theories are constructed with $\mathbb{Z}_{n}$ orbifolds on the internal $\mathbb{C}^{2}$. They were introduced in~\cite{Bergman:2012kr}, and the precise match of the large $N$ free energy with the holographic entanglement entropy was subsequently performed in~\cite{Jafferis:2012iv}. Here, we shall further match the central charges appearing on both sides of the holographic correspondence.

This section introduces these (infrared) quiver theories, and computes the large $N$ limit of their conformal and flavor central charges. In the next section, we match these central charges with the corresponding supergravity.

\subsection{Orbifold theories}

The orbifold superconformal field theories are constructed by taking the type I' string theory setup for the Seiberg theories as in Table~\ref{table:BraneDirections}, but replacing the flat $\mathbb{C}^{2}$ directions, spanned by $z_1 = x^{5}+ \ii x^{6}$ and $z_2= x^{7}+\ii x^{8}$, with a $\mathbb{C}^{2}/\mathbb{Z}_{n}$ orbifold, such that the D4-branes probe an ALE singularity.
For simplicity, let us begin with the brane configuration without the D8/O8.  At the orbifold fixed point, extra D6-branes can wrap the $n-1$ (vanishing) two-cycles $\Sigma_i$, $i = 1, \dotsc, n-1$, while being extended along the same 01234 directions as the D4-branes. A D6-brane wrapping $\Sigma_i$ carries $i$ units of the $\widetilde\bZ_n$ charge (labeling the twisted sectors), and is called a ``fractional'' D4-brane~\cite{Douglas:1996sw,Johnson:1996py}. In contrast, a D4-brane is uncharged under the $\bZ_n$.

While the type IIA frame is useful for deducing the supergravity dual, the infrared gauge theory is most naturally described in the type IIB frame, obtained by compactifying and T-dualizing the $x^5$ direction.\footnote{The type IIB brane webs were discussed in more details in~\cite{Bergman:2015dpa}.
}
There are $n$ NS5 branes located at $n$ points on the dual circle direction ${x'}^5$ and extended along the 012349 directions, with D5-branes stretched between. The low energy limit of the worldvolume theory of the D5-branes is described by a {\it cyclic} quiver gauge theory. Each D5-brane segment corresponds to an SU gauge node.

Next, let us consider the orientifold projection by the O7-plane, which acts simultaneously as worldsheet parity, and reflection on $x^9 \to -x^9$ as well as the dual circle, ${x'}^5 \to -{x'}^5$. Let us label the segments of the D5-branes by $a=1,\cdots,n$. For even $n=2k$, there are two different projections. The first possibility identifies the $a$-th D5-brane with the $(n-a)$-th D5-brane, for $a \neq k,n$, and projects the gauge group on the $k$-th and $n$-th D5-branes to USp, giving the {\it linear} quiver gauge theory in Figure~\ref{Fig:EvenVS}. The second identifies the $a$-th D5-brane with the $(n+1-a)$-th D5-brane, giving the quiver gauge theory in Figure~\ref{Fig:EvenNVS}. For odd $n=2k+1$, the orientifold projection identifies the $a$-th D5-brane with the $(n+1-a)$-th D5-brane, for $a \neq k+1$, and projects the gauge group on the $(k+1)$-th D5-brane to USp, giving the quiver gauge theory in Figure~\ref{Fig:Odd}.

Finally, we can spice up the above with flavors by adding $N_{\rm f}$ D8-branes to the type I' setup.  Depending on which of the three cases of orientifold-projected quivers is considered, different nonabelian flavor symmetries arise. We presently review the explicit forms of the (flavored) infrared quiver gauge theories and their global symmetries.

Let us remark here that the orbifold theories we discuss in the following exhibit intricate dualities and (possible) flavor symmetry enhancements.\footnote{See for instance~\cite{Bergman:2013aca,Zafrir:2014ywa}, where the $n=2$ theories at low ranks have been studied explicitly, and their dualities and flavor symmetry enhancement have been checked via the superconformal index. The authors further postulated conjectures for the higher rank cases, with supporting arguments based on five-dimensional web diagrams and their S-duality moves. We thank O. Bergman and D. Rodr{\'\i}guez-G{\'o}mez for correspondence on this point.} When computing the flavor central charges, we shall solely focus on the manifest flavor symmetry groups of the infrared gauge theories. It would be interesting to employ the techniques of~\cite{Chang:2017cdx} and the present paper to check the proposed dualities and flavor symmetry enhancements, as well as discover new ones, say, in the large $N$ limit using matrix model techniques.

\begin{figure}[H]
\centering
\includegraphics[width=0.70\textwidth]{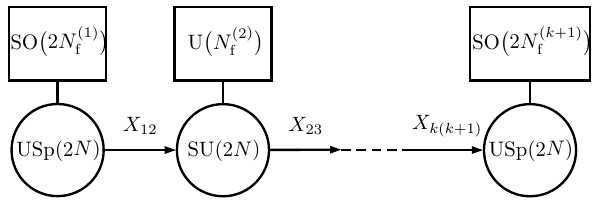}
\caption{Five-dimensional infrared quiver gauge theories giving rise to orbifold superconformal field theories with even $n=2k$ and with vector structure.}\label{Fig:EvenVS}
\end{figure}

\subsubsection*{Even orbifold with vector structure  (Figure~\ref{Fig:EvenVS})}

The gauge group of the infrared quiver theory is ${\rm USp}(2N) \times {\rm SU}(2N)^{k-1} \times {\rm USp}(2N)$, with $k$ bifundamental hypermultiplets $X_{a (a+1)}$ between adjoining gauge nodes. Each simple factor of the gauge group gives rise to a ${\rm U}(1)_{\rm I}$ instantonic symmetry, and thus, the overall instantonic symmetry group is ${\rm U}(1)_{\rm I}^{k+1}$. There is also a ${\rm U}(1)_{\rm b}^{(a)}$ symmetry for each bifundamental hypermultiplet $X_{a(a+1)}$. Finally, on each node, one can add $N_{\rm f}^{(a)}$ fundamental hypermultiplets, $a=1, \ldots, k+1$. The hypermultiplets coupled to the $a$-th ${\rm SU}(2N)$ factor, for $a = 2, \dotsc, k$, transform under ${\rm U}(N_{\rm f}^{(a)})$ flavor symmetry.  The ones coupled to each
the $a$-th ${\rm USp}(2N)$ factor, $a=1,k+1$, transform under manifest ${\rm SO}(2N_{\rm f}^{(a)})$ flavor symmetry. Overall, this leads to the following flavor symmetry group of the infrared gauge theory,
\ie
\hspace{-.15in} G_{\rm f}^{(n=2k), \, {\rm vs}}  \ &= \  {\rm U}(1)_{\rm I}^{(1)}\times {\rm U}(1)_{\rm b}^{(1)} \times {\rm SO}(2 N_{\rm f}^{(1)})  \times \prod_{a=2}^{k}\left[ {\rm U}(1)^{(a)}_{\rm I} \times {\rm U}(1)^{(a)}_{\rm b}\times  {\rm U}(N_{\rm f}^{(a)})\right] \\
 & \hspace{.5in} \times {\rm U}(1)_{\rm I}^{(k+1)} \times {\rm SO}(2 N_{\rm f}^{(k+1)}) \,.
\fe
The ${\rm U}(1)_{\rm b}^{(a)}$ factors can be recombined into a mesonic (diagonal) ${\rm U}(1)_{\rm M}$ symmetry, and $k-1$ baryonic ${\rm U}(1)_{\rm B}^{(a)}$ symmetries, $a=1, \ldots, k-1$. Their charges are related by
\ie\label{chargeRelation1}
Q_{\rm M} \ &= \ {1\over 2}\sum^{k}_{a=1}Q_{\rm b}^{(a)}\,,\\
Q_{\rm B}^{(a)} \ &= \ Q_{\rm b}^{(a)}-Q_{\rm b}^{(k)}\,.
\fe
The reason for this change of basis is such that the gauge-invariant meson and baryon operators span the charge lattice. The meson operator is
\ie
\label{meson1}
M  \ \equiv \ \tr \left[\prod_{a=1}^{k} X_{a (a+1)} \right]^{2} \,,
\fe
and the baryon operators are
\ie
B_{a}  \ = \ \det \left(X_{a(a+1)} \right) \,, \quad a = 1, \ldots , k \,.
\fe

\begin{figure}[H]
\centering
\includegraphics[width=.85\textwidth]{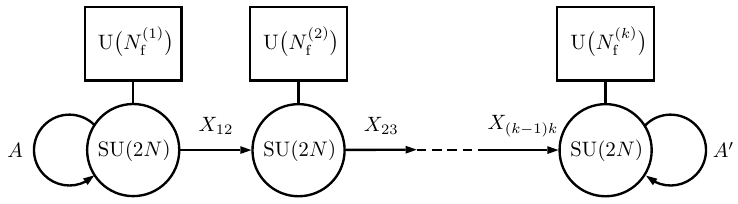}
\caption{Five-dimensional infrared quiver gauge theories giving rise to orbifold superconformal field theories with even $n=2k$ and without vector structure.}\label{Fig:EvenNVS}
\end{figure}

\subsubsection*{Even orbifold without vector structure (Figure~\ref{Fig:EvenNVS})}

The gauge group of the infrared quiver theory is ${\rm SU}(N)^{k}$, with bifundamental hypermultiplets $X_{a (a+1)}$ ($a=1, \ldots, k-1$) between adjoining gauge nodes, as well as two antisymmetric hypermultiplets $A$ and $A^{\prime}$ gauged separately by the first and the last nodes. For each node, $a=1, \ldots, k+1$, we add $N_{\rm f}^{(a)}$ fundamental hypermultiplets, each transforming under ${\rm U}(N_{\rm f}^{(a)})$ flavor symmetry.  There is an instantonic ${\rm U}(1)_{\rm I}^{(a)}$ symmetry for each gauge node, a ${\rm U}(1)_{\rm b}^{(a)}$ symmetry for each bifundamental hypermultiplet $X_{a (a+1)}$, and finally, ${\rm U}(1)_{\rm A}$, ${\rm U}(1)_{\rm A'}$ for the antisymmetric hypermultiplets $A$, $A'$. Together, we obtain the following global symmetry group of the infrared quiver gauge theory,
\ie
G_{\rm f}^{(n=2k), \, {\rm nvs}} \ &= \ {\rm U}(1)_{\rm A} \times \prod_{a=1}^{k-1}\left[  {\rm U}(1)^{(a)}_{\rm I} \times {\rm U}(1)^{(a)}_{\rm b}\times {\rm U}(N_{\rm f}^{(a)})\right]
\\
& \hspace{.5in} \times {\rm U}(1)^{(k)}_{\rm I}\times {\rm U}(1)_{\rm A'}\times {\rm U}(N_{\rm f}^{(k)})\,.
\fe
We can group the ${\rm U}(1)$ global symmetries coming from the bifundamental and the antisymmetric hypermultiplets into mesonic ${\rm U}(1)_{\rm M}$ and baryonic ${\rm U}(1)_{\rm B}^{(a)}$ factors, with $a = 1, \dotsc, k$.  Their charges are related by
\ie\label{chargeRelation2}
Q_{\rm M} \ &= \ {1\over 2}(Q_{\rm A}+Q_{\rm A'})+{1\over 2}\sum^{k-1}_{a=1}Q_{\rm b}^{(a)}\,,
\\
Q_{\rm B}^{(a)} \ &= \ Q_{\rm b}^{(a)}-Q_{\rm A}-Q_{\rm A'}\,,
\\
Q_{\rm B}^{(k)} \ &= \ Q_{\rm A}-Q_{\rm A'}\,.
\fe
The meson operator is
\ie
\label{meson2}
M  \ \equiv \ \tr \left[A \left( \prod_{a=1}^{k-1} X_{a (a+1)}^{2} \right) A' \right] \,.
\fe
A basis for the baryon operators is
\ie
B_{a}  \ = \ \det \left(X_{a(a+1)} \right) \,, \quad a = 1, \ldots , k-1 \,,
\fe
plus an additional one given by the Pfaffian of the antisymmetric hypermultiplet $A$,
\ie
B_{k} \ = \ {\rm Pf} (A) \ = \ \epsilon^{\alpha_1 \cdots \alpha_{2N}} A_{\alpha_1 \alpha_2} \cdots A_{\alpha_{2N-1} \alpha_{2N}} \,,
\fe
where $\alpha_j$ are the ${\rm USp}(2N)$-indices, raised and lowered by the corresponding invariant tensor. Note that the alternative combination ${\rm Pf} (A')$ is not independent of the chosen basis $\left\{ M, B_{1}, \ldots , B_{k} \right\}$ of gauge-invariant operators composed purely of hypermultiplets.

\begin{figure}[H]
\centering
\includegraphics[width=.775\textwidth]{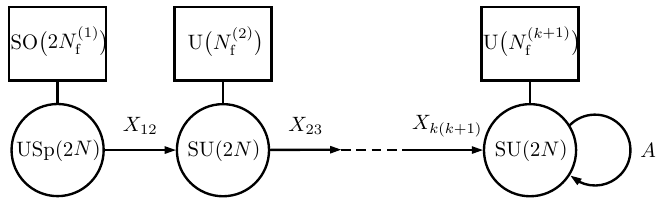}
\caption{Five-dimensional infrared quiver gauge theories giving rise to orbifold superconformal field theories with odd $n=2k+1$.}\label{Fig:Odd}
\end{figure}

\subsubsection*{Odd orbifold (Figure~\ref{Fig:Odd})}

This final case is a hybrid between the two even cases:  there is a ${\rm USp}(2N) \times {\rm SU}(2N)^{k}$ infrared gauge group, together with fundamental, bifundamental, and antisymmetric hypermultiplets.  The resulting overall infrared flavor symmetry group is
\ie
G_{\rm f}^{(n=2k+1)} \ &= \ {\rm U}(1)_{\rm I}^{(1)} \times {\rm U}(1)_{\rm b}^{(1)} \times {\rm SO}(2 N_{\rm f}^{(1)}) \times \prod_{a=2}^{k}\left[  {\rm U}(1)^{(a)}_{\rm I} \times {\rm U}(1)^{(a-1)}_{\rm b}\times {\rm U}(N_{\rm f}^{(a)})\right]
\\
& \hspace{.5in} \times {\rm U}(1)^{(k+1)}_{\rm I}\times {\rm U}(1)_{\rm A} \times  {\rm U}(N_{\rm f}^{(k+1)})\,.
\fe
As before, we group the ${\rm U}(1)$ symmetries into mesonic U(1)$_{\rm M}$ and baryonic U(1)$_{\rm B}^{(a)}$ factors, with $a=1,\cdots,k$. Their charges are related by 
\ie\label{chargeRelation3}
&Q_{\rm M} \ = \ {1\over 2}Q_{\rm A}+{1\over 2}\sum^k_{a=1}Q_{\rm b}^{(a)} \,,
\\
&Q_{\rm B}^{(a)} \ = \ Q_{\rm b}^{(a)}-2Q_{\rm A}\,.
\fe
The meson operator is
\ie
\label{meson3}
M  \ \equiv \ \tr \left[\left( \prod_{a=1}^{k} X_{a (a+1)}^{2}  \right)A \right] \,.
\fe
A basis for the baryon operators is
\ie
B_{a}  \ = \ \det \left(X_{a(a+1)} \right) \,, \quad a = 1, \ldots , k \,.
\fe
Again, the Pfaffian ${\rm Pf}(A)$ is not independent of the basis $\left\{ M, B_{1}, \ldots , B_{k} \right\}$ of gauge-invariant operators composed purely of hypermultiplets.

\vspace{.2in}

In all three cases, the total number of fundamental hypermultiplets equals the number of D8-branes in the type I' string theory setup, {\it i.e.}, $N_{\rm f} = \sum_a N_{\rm f}^{(a)}$. This number must fall into the range $0 \leq N_{\rm f} < 8$.

\subsection{Free energy and conformal central charge}

We begin by writing down the localized action of the Coulomb branch integral expression~\eqref{spf} for the perturbative five-sphere partition function. For ease of notation, it is convenient to introduce a set of auxiliary Coulomb branch parameters $\mu_i^{(a)}$, where $a$ labels the gauge node, $i = 1, \dotsc, 2N$, and then perform identifications on $\mu_i^{(a)}$ suitable for the specific gauge group.  Schematically, the Coulomb branch integration measure is
\ie
\int [d\mu] \, e^{-F(\mu)} \, \delta(\text{constraints})\,,
\fe 
up to some ambiguity in the Jacobian factor that only affects the free energy at order $N$ in the large $N$ limit.  Since we only need to be accurate to order $N^{3/2}$
to extract the central charges of interest, we shall be liberal in our definition of this integration measure.  The forms of the localized action $F(\mu)$ for the three classes of orbifold theories are as follows.

\subsubsection*{Even orbifold with vector structure}
\ie\label{evenFvs}
F^{(2k)}_{{\rm vs}}(\mu) \ &= \ \frac{1}{2} \sum^{2N}_{i \neq j} \Bigg[ \frac{1}{2} G_{V} \left(\pm (\mu_{i}^{(1)} -\mu_{j}^{(1)}) \right) + \sum_{a=2}^{k} G_{V} \left( \pm(\mu_{i}^{(a)} -\mu_{j}^{(a)} )\right) \\
&\hspace{.5in} 
+ \frac{1}{2} G_{V} \left( \pm(\mu_{i}^{(k+1)} -\mu_{j}^{(k+1)}) \right)  \Bigg]
+\sum^{2N}_{i,j=1} \left[ \sum_{a=1}^{k} G_{H} \left(\mu_{i}^{(a)} -   \mu_{j}^{(a+1)}  \right) \right] \\
& \ \quad
+ \sum^{2N}_{i=1} \left[\frac{1}{2} G_{V}\left(2\mu_{i}^{(1)}\right) + \frac{1}{2} G_{V}\left(2\mu_{i}^{(k+1)}\right) + \sum_{a=1}^{k+1} N_{\rm f}^{(a)} G_{H}\left(\mu_{i}^{(a)} \right)\right] \,,
\fe
with the constraints $\mu_{N+i}^{(1)}= - \mu_{i}^{(1)}$, $\mu_{N+i}^{(k+1)} = - \mu_{i}^{(k+1)}$, and $\sum_{i=1}^{2N} \mu_{i}^{(a)}=0$ for $2\leq a \leq k$.

\subsubsection*{Even orbifold without vector structure}
\ie\label{evenFnvs}
\hspace{-.15in} F^{(2k)}_{{\rm nvs}}(\mu) \ &= \ 
 \frac{1}{2} \sum^{2N}_{i \neq j} \Bigg[ \sum_{a=1}^{k} G_{V} \left(\pm (\mu_{i}^{(1)}-\mu_{j}^{(1)}) \right) 
+ G_{H} \left(\mu_{i}^{(1)} +\mu_{j}^{(1)} \right)
+ G_{H} \left(\mu_{i}^{(k)} +\mu_{j}^{(k)} \right) \Bigg] \\
& \ \quad 
+\sum^{2N}_{i,j=1} \left[ \sum_{a=1}^{k-1} G_{H} \left(\mu_{i}^{(a)} -   \mu_{j}^{(a+1)}  \right) \right] 
+ \sum_{i=1}^{2N} \left[ \sum_{a=1}^{k} N_{\rm f}^{(a)} G_{H}\left(\mu_{i}^{(a)}\right) \right] \,,
\fe
with $\sum_{i=1}^{2N} \mu_{i=1}^{(a)}=0$ for $1\leq a \leq k$. 

\subsubsection*{Odd orbifold}
\ie\label{oddF}
F^{(2k+1)}(\mu) \ &= \ 
\frac{1}{2} \sum^{2N}_{i \neq j} \Bigg[ \frac{1}{2} G_{V} \left(\pm (\mu_{i}^{(1)} -\mu_{j}^{(1)}) \right) 
+ \sum_{a=2}^{k+1} G_{V} \left(\pm(\mu_{i}^{(a)} -\mu_{j}^{(a)} )\right) \\
&\hspace{.5in} 
+G_{H} \left(\mu_{i}^{(k+1)} +\mu_{j}^{(k+1)} \right)  \Bigg]
+\sum^{2N}_{i,j} \left[ \sum_{a=1}^{k} G_{H} \left(\mu_{i}^{(a)} -   \mu_{j}^{(a+1)}  \right) \right] \\
&\ \quad
+ \sum^{2N}_{i=1} \left[\frac{1}{2} G_{V}\left(2\mu_{i}^{(1)}\right) 
+ \sum_{a=1}^{k+1} N_{\rm f}^{(a)} G_{H}\left(\mu_{i}^{(a)} \right)\right] \,,
\fe
with $\mu_{N+i}^{(1)} = - \mu_{i}^{(1)}$ and $\sum_{i=1}^{2N} \mu_{i}^{(a)} = 0$.

\hspace{.2in}

In the large $N$ limit, provided that $\mu_{i}^{(a)}$ scale as $N^{1/2}$, the leading terms in the above $F(\mu)$ scale as $N^{7/2}$, and are minimized by~\cite{Jafferis:2012iv}
\ie
 \mu_{i}^{(a)} \ &= \ \mu_{i} \,,\quad && 1\leq i \leq 2N\,,\label{mulambdaii}\\
 \mu_{i}  \ = \ - \mu_{N+i}\ &\equiv \ \lambda_i \,,\quad && 1 \leq i \leq N \,,
\fe
where $\lambda_i$ become the sole remaining Coulomb branch parameters. As before, define
\ie
\lambda_i \ = \ N^{1/2} x_i\,,
\fe
and rewrite the localized action $F(\mu)$ as a functional $F[\rho]$ of the density $\rho(x)$.
The equations \eqref{mulambdaii} actually render the ${\rm SU}(2N)$ and the ${\rm USp}(2N)$ gauge factors indistinguishable. Now we can take the large $N$ limit of the exponent, and find that for all three cases, it is simply given by
\ie
\label{Eqn:Forbfsq}
\hspace{-.15in} F[\rho] &= - {N^{5/2} \over \omega_1 \omega_2 \omega_3} \int_{0}^{x_\star} \diff x \, \rho(x) \int_{0}^{x_\star}  \diff y \, \rho(y)  \left[ \frac{n \pi \omega_{\mathrm{tot}}^{2}}{8} \left( x+y+|x-y| \right)  - \frac{(8-N_{\rm f}) \pi}{3} x^{3}\right]
\\
 & \hspace{1.5in} + \mathcal{O}( N^{3/2} ) \, .
\fe
The large $N$ localized action $F[\rho]$ receives two types of corrections. First, there are $N^{3/2}$ terms coming from the asymptotic expansions \eqref{expansionhyper} of $F(\m)$. Second, there are order $N^{-1}$ corrections to the saddle point configuration \eqref{mulambdaii} due to the $N^{5/2}$ terms in $F(\m)$, and they also give rise to order $N^{7/2} \times (N^{-1})^2 = N^{3/2}$ corrections to $F[\rho]$.\footnote{A correction $\delta\mu$ to the saddle point configuration \eqref{mulambdaii} results in a correction of order $\delta\mu^2$ to the localized action $F(\mu)$, since by definition, $\delta F(\mu)/\delta\mu = 0$ at the saddle point.
}
The saddle point approximation to the Coulomb branch integral can be performed in complete analogy to the case of Seiberg theories, leading to the saddle point configuration
\ie
\label{SaddleOrbifold}
\rho(x) \ = \ \frac{x}{2x_\star^2} \,,\qquad
x_{\star}^2 \ = \ \frac{9n}{2(8-{N_{\rm f}})} \,,
\fe
and the free energy
\bea
\label{Fn}
F \ = \  - \frac{\sqrt{2} \,  \pi \omega_{\mathrm{tot}}^{3}}{15 \, \omega_1\omega_2\omega_3} {n^{3/2}N^{5/2} \over \sqrt{8-N_{\rm f}}} + \mathcal{O}( N^{3/2}) \,.
\eea
Invoking the general relation between the squashed free energy and the conformal central charge~\eqref{CTRelation}, we find that the conformal central charges of the family of orbifold theories are given by 
\ie
\label{CTlargeNft}
C_T \ &= \  \frac{1152 \sqrt{2}}{\pi} {n^{3/2}N^{5/2} \over \sqrt{8-N_{\rm f}}} + {\cal O}(N^{3/2}) \,.
\fe

\subsection{Flavor central charges}

Let us now turn to the computation of the flavor central charges for the orbifold theories.  It suffices to consider the round sphere, so we set $\omega_i = 1$ in the following.

\subsubsection{Fundamental hypermultiplet symmetries}

We first compute the flavor central charges for generic ${\rm SO}(2N_{\rm f}^{(a)})$ factors -- associated with the gauge nodes $a=1$ and $a=k+1$ for even orbifold with vector structure, and $a=1$ for odd orbifold -- in the flavor symmetry groups of the orbifold theories. The modification to the localized action $F(\mu)$ due to the introduction of a mass matrix
\ie
 \cM({\rm SO}(2N_{\rm f}^{(a)})) \ = \ m_\mathrm{f}^{(a)} \,\left( \ii \sigma_2 \otimes \mathbbm{1}_{N_{\rm f}} \right) \ \in \ \mathfrak{so}( 2N_{\rm f}^{(a)} ) \,
\fe
is given by
\ie
\label{ExpoCorrection}
\Delta F(\mu) \ &= \  N_{\rm f}^{(a)} \sum_{i=1}^{2N}\left[ G_{H}\left(\mu_{i}^{(a)} + m_{\rm f}^{(a)} \right) - G_{H}\left(\mu_{i}^{(a)} \right) \right] \,.
\fe
The leading order correction to the free energy is given by the large $N$ asymptotics \eqref{expansionhyper} of $\Delta F(\mu)$ evaluated on the leading order saddle~\eqref{mulambdaii} as well as \eqref{SaddleOrbifold} (from extremizing the leading order $N^{7/2}$ piece of the exponent),
\ie
\Delta F \ = \ \frac{\sqrt{2} \, \pi N_{\rm f}^{(a)}  (m_{\rm f}^{(a)})^2}{\sqrt{8-{N_{\rm f}}}} n^{1/2} N^{3/2} + \cO(N^{1/2}) \,.
\fe
Using
\ie
{I}_{{\bf vec}({\rm SO}(2N_{\rm f}))} \ = \ 1, \qquad \tr_{{\bf vec}({\rm SO}(2N_{\rm f}))}({\cal M}^2) 
\ = \ 2 N_{\rm f}^{(a)} (m_{\rm f}^{(a)})^2,
\fe
in the relation~\eqref{CJRelation2}, we determine the corresponding flavor symmetry central charges to be
\ie\label{Eqn:CJSO2NfOrb}
C_{J}^{{\rm SO}(2N_{\rm f}^{(a)})} \ = \  \frac{256 \sqrt{2}}{3 \pi\sqrt{8-N_{\rm f}}} \, n^{1/2} N^{3/2} + \cO(N^{1/2})\,.
\fe

The fact that the leading $1/N$ corrections to the flavor central charges are of order $N^{1/2}$ needs some explanation.  Firstly, one may worry that the order $N^{3/2}$ corrections to the undeformed free energy $F$ in \eqref{Fn} would contaminate the flavor central charges that are also of order $N^{3/2}$.  However, these corrections do not depend on the mass parameters, and therefore do not affect the flavor central charges.
Secondly, since the mass deformations given by \eqref{ExpoCorrection} scale as $m^2 N^{3/2}$ ($N^{-2}$ relative to the leading $N^{7/2}$), there are corrections to the saddle point configuration \eqref{mulambdaii} that are of order $m^2 N^{-2}$, in addition to the $m$-independent corrections of order $N^{-1}$ discussed below \eqref{Fn}. The corresponding correction to the free energy $F$ in \eqref{Fn} that is quadratic in $m$ is from the mixed term, of order $N^{7/2} \times (m^2N^{-2}) \times N^{-1} = m^2 N^{1/2}$.

We now consider the remaining gauge nodes that are associated with
 ${\rm U}(N_{\rm f}^{(a)})$ flavor factors, by introducing the mass matrix,
\ie
 \cM({\rm U}(N_{\rm f}^{(a)})) \ = \ \ii  m_\mathrm{f}^{(a)} \, \mathbbm{1}_{N_{\rm f}^{(a)}} \ \in \ \mathfrak{u}( N_{\rm f}^{(a)}) \,,
\fe
into the Lagrangian.  Proceeding as before, we find that the leading order correction to the free energy is
\ie
\Delta F \ = \ \frac{\sqrt{2}\,\pi N_{\rm f}^{(a)}  (m_{\rm f}^{(a)})^2}{\sqrt{8-{N_{\rm f}}}} n^{1/2} N^{3/2} + \cO(N^{1/2}) \,.
\fe
Using
\ie
I_{\mathbf{fund} ({\rm U}(N_{\rm f}^{(a)}))} \ = \ \frac{1}{2} \,, \qquad \tr_{\mathbf{fund} ({\rm U}(N_{\rm f}^{(a)}))} (\cM^{2}) \ = \ N_{\rm f}^{(a)} (m_{\rm f}^{(a)})^{2} \,
\fe
in the relation~\eqref{CJRelation2}, we determine the corresponding flavor symmetry central charges to be
\ie\label{Eqn:CJUNfOrb}
C_{J}^{{\rm U}(N_{\rm f}^{(a)})} \ = \  \frac{256 \sqrt{2}}{3 \pi\sqrt{8-N_{\rm f}}} \, n^{1/2} N^{3/2} + \cO(N^{1/2})\,.
\fe

As will be discussed further in Section~\ref{Sec:ComparisonToSeiberg}, Hanany-Witten moves~\cite{Hanany:1996ie} relate an orbifold theory considered here to one that has ${\rm SO}(2N_{\rm f})$ flavor acting on hypermultiplets charged under the first gauge node, and the finite shifts in the ranks of the gauge nodes cannot be detected in the large $N$ limit. This large $N$ flavor symmetry enhancement explains the coincidence of the flavor central charges for all gauge nodes in the original orbifold theory.

\subsubsection{Bifundamental hypermultiplet symmetries}

We now turn towards the flavor central charges for the U(1)$^{(a)}_{\rm b}$ flavor symmetry factors. Each bifundamental hypermultiplet $X_{a(a+1)}$ has unit charge under the U(1)$^{(a)}_{\rm b}$ flavor symmetry associated to a mass parameter $m_{\rm b}^{(a)}$ for $X_{a(a+1)}$. The mass deformation of the localized action is 
\ie
\Delta F(\mu) \ &= \
 \sum^{2N}_{i,j=1} \left[  G_{H} \left(\mu_{i}^{(a)} -   \mu_{j}^{(a+1)} + 2m_{\rm b}^{(a)}  \right) - G_{H} \left(\mu_{i}^{(a)} -   \mu_{j}^{(a+1)} \right)  \right] \,.
\fe
Once again, evaluating $\Delta F(\mu)$ on the large $N$ saddle~\eqref{mulambdaii}, and taking the continuum limit, the localized action becomes
\ie
\Delta F[\rho] \ = \ - 4\pi \big(m_{\rm b}^{(a)}\big)^{2} N^{5/2} \int_{0}^{x_{\star}} \diff x\,\rho (x) \int_{0}^{x_{\star}} \diff y\,\rho(y) \left(  x+y + \left| x-y \right| \right) + \cO(N^{3/2}) \,.
\fe
This term modifies the saddle to
\ie
\rho(x) \ = \ \frac{2x}{x_\star^2} \,,\qquad
x_{\star}^2 \ = \ \frac{ 9n + 32 (m_{\rm b}^{(a)})^2}{2(8-{N_{\rm f}})} \,,
\fe
which leads to the large $N$ free energy,
\ie 
F \ = \ - \frac{\sqrt{2} \pi \left(9n + 32 (m_{\rm b}^{(a)})^2\right){}^{3/2}}{15 \sqrt{8-{N_{\rm f}}}}\,   N^{5/2} + {\cal O}(N^{3/2})\,.
\fe
By the relation \eqref{CJRelation1}, we are led to the following U(1)$^{(a)}_{\rm b}$ flavor central charge
\ie\label{Eqn:CJU1bOrb}
C_{J}^{{\rm U}(1)_{\rm b}} \ = \ \frac{2048 \sqrt{2}}{5\pi \sqrt{8-N_{\rm f}}} \, n^{1/2} N^{5/2} + {\cal O}(N^{3/2}) \,.
\fe

\subsubsection{Antisymmetric hypermultiplet symmetries}
The antisymmetric hypermultiplet $A$ in the odd orbifold has unit charge under the U(1)$_{\rm A}$ flavor symmetry associated with a mass parameter $m_{\rm A}$. The mass deformation of the localized action is
\ie
\Delta F(\mu) \ &= \ 
\frac{1}{2} \sum^{2N}_{i \neq j}\left[G_{H} \left(\mu_{i}^{(k+1)} +\mu_{j}^{(k+1)} + 2 m_{\rm A}\right) - G_{H} \left(\mu_{i}^{(k+1)} +\mu_{j}^{(k+1)} \right)\right] \,.
\fe
Evaluating $\Delta F(\mu)$ on the large $N$ saddle~\eqref{mulambdaii}, and taking the continuum limit, the localized action becomes
\ie
\Delta F[\rho] \ = \ - 2\pi \big(m_{\rm A}\big)^{2} N^{5/2} \int_{0}^{x_{\star}} \diff x\,\rho (x) \int_{0}^{x_{\star}} \diff y\,\rho(y) \left(  x+y + \left| x-y \right| \right) + \cO(N^{3/2}) \,.
\fe
This term modifies the saddle to
\ie
\rho(x) \ = \ \frac{2x}{x_\star^2} \,,\qquad
x_{\star}^2 \ = \ \frac{9n + 16 m_{\rm A}^2}{2(8-{N_{\rm f}})} \,,
\fe
which gives the large $N$ free energy,
\ie
F \ = \ - \frac{\sqrt{2} \pi \left(9n + 16 m_{\rm A}^2\right){}^{3/2}}{15 \sqrt{8-{N_{\rm f}}}}\,   N^{5/2} + {\cal O}(N^{3/2})\,.
\fe
By the relation \eqref{CJRelation1}, we are led to the following ${\rm U(1)}^{(a)}_{\rm A}$ flavor central charge
\ie\label{Eqn:CJU1AOrb}
C_{J}^{{\rm U}(1)_{\rm A}} \ = \ \frac{1024 \sqrt{2}}{5\pi \sqrt{8-N_{\rm f}}} \, n^{1/2} N^{5/2} + {\cal O}(N^{3/2}) \,.
\fe
An analogue analysis shows that the antisymmetric hypermultiplets $A$ and $A'$ in the even orbifold without vector structure have the same flavor central charge as in \eqref{Eqn:CJU1AOrb}.

\subsubsection{Mesonic and baryonic symmetries}

As specified in~\eqref{chargeRelation1},~\eqref{chargeRelation2} and~\eqref{chargeRelation3}, the mesonic ${\rm U(1)}_{\rm M}$ and baryonic ${\rm U(1)}_{\rm B}$ symmetries are linear combinations of the U(1) flavor symmetry factors that act on the bifundamental and antisymmetric hypermultiplets. The mesonic flavor central charge is then given by
\ie\label{Eqn:CJU1MOrb}
C_J^{{\rm U(1)}_{\rm M}} \ = \ {n\over 8}C_{J}^{{\rm U}(1)_{\rm b}} \ = \ \frac{256 \sqrt{2}}{5\pi \sqrt{8-N_{\rm f}}} \, n^{3/2} N^{5/2} + {\cal O}(N^{3/2}) \,.
\fe

The baryonic symmetries are not orthogonal. Their flavor central charges are given by matrices. These (symmetric) matrices can be computed by taking derivatives with respect to the mass parameters of different baryonic U(1) factors in the overall flavor symmetry. We write these matrices in the bases \eqref{chargeRelation1}, \eqref{chargeRelation2}, and \eqref{chargeRelation3}. In even orbifolds with vector structure, the baryonic flavor central charge matrix is
\ie\label{CJU1B1}
C_J^{{\rm U(1)}_{\rm B}} \ &= \ \begin{pmatrix}
2 \ & \ 1 \ & \ 1 \ & \cdots & \ 1 \ \\
1 \ & \ 2 \ & \ 1 \ & \cdots & \ 1\\
  &  & \vdots &  &  \\
1 \ & \ 1 \ & \cdots & \ 1 \ & \  2 \\
\end{pmatrix}_{(k-1) \times (k-1)} \hspace{-.5in} 
\times C_{J}^{{\rm U}(1)_{\rm b}}  \,.
\fe
In even orbifolds without vector structure,
\ie\label{CJU1B2}
C_J^{{\rm U(1)}_{\rm B}} \ &= \ \begin{pmatrix}
 2 \ & \ 1 \ & \ 1 \ & \cdots & \ 1 \ & \ 0 \ \\
 1 \ & \ 2 \ & \ 1 \ & \cdots & \ 1 \ & \ 0 \ \\
  &  & \vdots &  &  \\
 1 \ & \ 1 \ & \cdots & \ 1 \ & \ 2 \ & \ 0 \ \\
 0 \ & \ 0 \ & \cdots & \ 0 \ & \ 0 \ & \ 1 \ \\
\end{pmatrix}_{k \times k}
\times C_{J}^{{\rm U}(1)_{\rm b}}  \,.
\fe
In odd orbifolds,
\ie\label{CJU1B3}
C_J^{{\rm U(1)}_{\rm B}} \ &= \ \begin{pmatrix}
 3 \ & \ 2 \ & \ 2 \ & \cdots & \ 2 \  \\
 2 \ & \ 3 \ & \ 2 \ & \cdots & \ 2\\
 &  & \vdots &  &  \\
 2 \ & \ 2 \ & \cdots & \ 2 \ & \ 3 \ \\
\end{pmatrix}_{k \times k}
\times C_{J}^{{\rm U}(1)_{\rm b}}  \,.
\fe

To compare with the results from gravity in Section \ref{sec:GravityBS}, it is convenient to perform a change of basis for the charge lattice from the $Q_{\rm B}^{(a)}$ defined in \eqref{chargeRelation1}, \eqref{chargeRelation2}, and \eqref{chargeRelation3}. For even orbifold with vector structure, we define
\ie
Q_{\Omega_a} \ &= \ Q_{\rm B}^{(a)}-Q_{\rm B}^{(a+1)} \quad {\rm for} \quad a=1,\cdots,k-2\,,
\\
Q_{\Omega_{k-1}} \ &= \ Q_{\rm B}^{(k-1)}\,.
\fe
For even orbifold without vector structure, we define
\ie
Q_{\Omega_a} \ &= \ Q_{\rm B}^{(a)}-Q_{\rm B}^{(a+1)}\quad{\rm for}\quad a=1,\cdots,k-2\,,
\\
Q_{\Omega_{k-1}} \ &= \ Q_{\rm B}^{(k-1)}\,,
\\
Q_{\Omega_{k}} \ &= \ Q_{\rm B}^{(k)}\,.
\fe
For odd orbifold without vector structure, we define
\ie
Q_{\Omega_a} \ &= \ Q_{\rm B}^{(a)}-Q_{\rm B}^{(a+1)}\quad{\rm for}\quad a=1,\cdots,k-1\,,
\\
Q_{\Omega_{k}} \ & = \ Q_{\rm B}^{(k)}\,.
\fe
In these bases, we find
\ie
C_{J,{\rm even, vs}}^{{\rm U(1)}_{\rm B}} \ &= \ \begin{pmatrix}
2 \ & \ -1 \ & \ 0 \ & \ 0 \ &\cdots & \ 0 \ & \ 0 \ \\
-1 \ & \ 2 \ & \ -1 \ & \ 0 \ &\cdots & \ 0 & \ 0 \ \\
 & & & \ \vdots \ &  &  \\
0 \ & \ 0 \ &\cdots & \ 0 \ & \ -1 \ & \ 2 \ & \ -1 \ \\
0 \ & \ 0 \ &\cdots & \ 0 \ & \ 0 \ & \ -1 \ & \  2 \\
\end{pmatrix}_{(k-1) \times (k-1)} \hspace{-.5in}
\times C_{J}^{{\rm U}(1)_{\rm b}}  \,,
\\
~
\\
C_{J,{\rm even, nvs}}^{{\rm U(1)}_{\rm B}} \ &= \ \begin{pmatrix}
2 \ & \ -1 \ & \ 0 \ & \ 0 \ &\cdots & \ 0 \ & \ 0 \ & \ 0 \ \\
-1 \ & \ 2 \ & \ -1 \ & \ 0 \ &\cdots & \ 0 & \ 0 \ & \ 0 \ \\
 & & & \ \vdots \ &  &  \\
0 \ & \ 0 \ &\cdots & \ 0 \ & \ -1 \ & \ 2 \ & \ -1 \ & \ 0 \ \\
0 \ & \ 0 \ &\cdots & \ 0 \ & \ 0 \ & \ -1 \ & \ 2 \ & \ 0 \ \\
0 \ & \ 0 \ &\cdots & \ 0 \ & \ 0 \ & \ 0 \ & \ 0 \ & \ 1 \ 
\end{pmatrix}_{k \times k}
\times C_{J}^{{\rm U}(1)_{\rm b}}  \,,
\\
~
\\
C_{J,{\rm odd}}^{{\rm U(1)}_{\rm B}} \ &= \ \begin{pmatrix}
2 \ & \ -1 \ & \ 0 \ & \ 0 \ &\cdots & \ 0 \ & \ 0 \ \\
-1 \ & \ 2 \ & \ -1 \ & \ 0 \ &\cdots & \ 0 & \ 0 \ \\
 & & & \ \vdots \ &  &  \\
0 \ & \ 0 \ &\cdots & \ 0 \ & \ -1 \ & \ 2 \ & \ -1 \ \\
0 \ & \ 0 \ &\cdots & \ 0 \ & \ 0 \ & \ -1 \ & \  3 \\
\end{pmatrix}_{k \times k}
\times C_{J}^{{\rm U}(1)_{\rm b}}  \,.
\label{CJBmatrix}
\fe
The above change of basis, \eqref{CJU1B1}, \eqref{CJU1B2}, and \eqref{CJU1B3}, from $Q_{\rm B}^{(a)}$ to $Q_{\Omega_a}$ is integral and uni-modular, and therefore preserves the charge lattice. While such a change of basis is convenient for the later holographic comparison, more fundamentally, we are matching the charge lattice, which is basis-independent.

\subsubsection{Instantonic symmetries}
\label{sec:instantonOrb}

In order to extract the flavor central charge for the ${\rm U}(1)_{\rm I}$ instantonic factor associated to the $a$-th gauge node, we proceed as in Section~\ref{Sec:FlavorU1J} for the Seiberg theories, by keeping the contribution of the classical piece. The large $N$ localized action is
\ie\label{Eqn:Forbfsq1}
 F[\rho] \ &= \
-\frac{9 n \pi}{8}   N^{5/2} \int_{x_1}^{x_2} \diff x  \, \rho(x)\int_{x_1}^{x_2} \diff y \, \rho(y) \left( x+y+|x-y| \right)
\\  
& \hspace{.5in} + \frac{(8-N_{\rm f}) \pi}{3} N^{5/2} \int_{x_1}^{x_2} \diff x \, \rho(x) x^{3} + {\cal O}(N^{3/2})
\\
& \hspace{.5in} + \sum_{a}m_{\rm I}^{(a)} \left[ 2\pi N^{2} \int_{x_1}^{x_2} \, \diff x \, \rho(x) x^2 + {\cal O}(N) \right] \, ,
\fe
for either the $\rm USp$ or $\rm SU$ gauge nodes.
The saddle point solution to the relevant $1/N$ order is
\ie
\rho(x) \ &= \ \frac{4 (8-{N_{\rm f}}) x}{9 n} \,,
\\
{x_1} \ &= \ 0 \,,
\\
{x_2} \ &= \  \frac{4  \sum_{a}{m_{\rm I}^{(a)}}+\sqrt{16 ( \sum_{a} m_{\rm I}^{(a)})^2+18 \, n N(8-N_{\rm f}) }}{2 \sqrt{N} (N_{\rm f}-8)}\,,
\fe
and gives rise to the large $N$ free energy, 
\ie
 F \ &= \ 
- \frac{9 \sqrt{2} \pi}{5 \sqrt{8-{N_{\rm f}}}} n^{3/2} N^{5/2} + {\cal O}(N^{3/2})
+ \sum_{a} m_{\rm I}^{(a)} \left[ \frac{9 \pi }{2(8-N_{\rm f})} n N^2 + {\cal O}(N) \right] 
\\
& \hspace{.5in} - \left( \sum_{a}m_{\rm I}^{(a)}\right)^2 \left[ \frac{4 \sqrt{2} \pi}{(8-N_{\rm f})^{3/2}} {n}^{1/2} N^{3/2} + {\cal O}(N^{1/2}) \right] + {\cal O}((m_{\rm I}^{(a)})^3) \,.
\fe
As in the case of Seiberg theories, the linear $m_{\rm I}^{(a)}$ piece would be inconsistent with conformal symmetry, and must be removed by a counter-term.  Using the relation \eqref{CJRelation1}, we find that the instanton symmetries are not orthogonal, and their flavor central charge matrix is
\ie\label{eqn:CJU1IOrb}
C_{J}^{{\rm U}(1)_{\rm I}} \ &= \ \frac{1024 \sqrt{2} }{3\pi (8-N_{\rm f})^{3/2}} n^{1/2}N^{3/2} \begin{pmatrix}
1 \ & \ \cdots \ & \ 1 \ \\
 & \ \vdots \ &  \\
1 \ & \ \cdots \ & \ 1 \ \\
\end{pmatrix} + {\cal O}(N^{1/2}) \, ,
\fe
whose dimensionality equals the number of gauge nodes. The reasoning for the leading $1/N$ corrections to the flavor central charges being of order $N^{1/2}$ proceeds as before. Since the mass deformations in \eqref{Eqn:Forbfsq1} scale as $m N^2$ ($N^{-3/2}$ relative to the leading $N^{7/2}$), there are corrections to the saddle point configuration \eqref{mulambdaii} that are of order $m N^{-3/2}$, in addition to the $m$-independent corrections of order $N^{-1}$ discussed below \eqref{Fn}. The correction to the free energy $F$ in \eqref{Fn} that is quadratic in $m$ is then of order $N^{7/2} \times (m N^{-3/2})^2 = m^2 N^{1/2}$.

At order $N^{3/2}$, the flavor central charge matrix is rank-one, meaning that we only observe \emph{one} independent combination of flavor central charges. In order to access the other independent flavor central charges that are of order $N^{1/2}$, we are required to carefully study the subleading contributions to the matrix models~\eqref{evenFvs},~\eqref{evenFnvs}, and \eqref{oddF}, and in particular modify the leading order saddle~\eqref{mulambdaii}.

\section{Central charges from the supergravity dual of orbifold theories}

Let us now study the massive IIA supergravity duals for the orbifold theories~\cite{Bergman:2012kr}, and subsequently match the various flavor central charges associated to the global symmetry group.

The holographic duals of the orbifold theories are the $\mathbb{Z}_{n}$ orbifolds of the geometry $\cM_{6} \times_{\rm w} {\rm HS}^{4}$ discussed in Section~\ref{Sec:MassiveIIASeiberg}. More precisely, for each S$^{3}$-slice at constant $\A$ of the hemisphere, we impose the identification $\theta_3 \to \theta_3 + 4\pi /n$ on the coordinate $\theta_3$ in \eqref{Eqn:S3coords}.  This corresponds to replacing the three-sphere with the lens space $L(n,1) = {\rm S}^{3}/\mathbb{Z}_{n}$. Notice that this action reduces the volume of the four-hemisphere by a factor of $n$,
\ie\label{Eqn:HS4Orb}
{\rm Vol}_{{\rm HS}^4/\mathbb{Z}_n} \ = \ \frac{{\rm Vol}_{{\rm HS}^4}}{n}\,,
\fe
and affects the effective six-dimensional couplings upon compactification. Due to the volume reduction, the $F_4$ flux is multiplied by a factor of $n$ to preserve the D4-brane charge quantization condition~\eqref{Eqn:D4branechargeSeiberg}. Hence, the supergravity solution of the orbifold theory is given by the substitution $N \to nN$ in the Seiberg theory background \eqref{eqn:10metric}, \eqref{eqn:solsIIAPhim}, \eqref{4Form}, and \eqref{L/lsinN}. Together these considerations immediately give us many of the central charges.

\subsection{Central charges by comparison to Seiberg theories}
\label{Sec:ComparisonToSeiberg}

Compared to the holographic duals of Seiberg theories, the free energy receives an extra factor of $n^{-1} \times n^{5/2} = n^{3/2}$, where the $n^{-1}$ comes from the reduction of the internal volume, and the $n^{5/2}$ comes from the $N \to nN$ shift due to the modified charge quantization. This precisely matches the large $N$ conformal central charge~\eqref{CTlargeNft} of the field theory. By the same argument, the ${\rm SO}(2N_{\rm f})$ and ${\rm U}(N_{\rm f})$ flavor central charges each receives a factor of $n^{1/2}$, matching the field theory results in~\eqref{Eqn:CJSO2NfOrb} and~\eqref{Eqn:CJUNfOrb}, respectively.

In the large $N$ limit, we are not able to distinguish between the distinct theories that arise from different distributions of the 
\ie\label{eqn:NfSNfa}
N_{\rm f} \ = \ \sum_{a} N_{\rm f}^{(a)} 
\fe
 fundamental hypermultiplets, for the following reason. On the gravity side, the orbifold solutions in~\cite{Bergman:2012kr} describe the configurations where all the $N_{\rm f}$ D8-branes sit at the first gauge node. However, starting with a different setup, in which the $N_{\rm f}$ D8-branes are distributed among different gauge nodes, and performing consecutive Hanany-Witten moves~\cite{Hanany:1996ie}, one ends up with a field theory description in which some of the ranks of the gauge nodes are shifted $N \to N + \ell$, where $\ell \leq N_{\rm f}$. But this effect is not expected to be visible in the leading order large $N$ asymptotics. Thus, in the strict large $N$ limit, we will only ever be able to probe the ${\rm SO}(2N_{\rm f})$ flavor symmetry associated to the final gauge node.\footnote{As a matter of fact, we cannot even distinguish between ${\rm SO}(2N_{\rm f})$ and ${\rm U}(N_{\rm f})$ by their flavor central charges at leading large $N$.
}
 
The mesonic ${\rm U}(1)_{\rm M}$ symmetry of the orbifold theories corresponds to the ${\rm U}(1)_{\rm M}$ factor in the isometry group of the orbifold hemisphere, ${\rm SU}(2)_{\rm R}\times {\rm U}(1)_{\rm M}$. Compared to the flavor central charge of the ${\rm U}(1)_{\rm M} \subset {\rm SU}(2)_{\rm M}$ subgroup in the Seiberg theories, the orbifold theories receive the same extra $n^{3/2}$ factor as discussed above for the conformal and hypermultiplet flavor central charges. Taking into account the embedding index
\ie
I_{\mathfrak{u}(1)_{\rm M} \hookrightarrow \mathfrak{su}(2)_{\rm M}} \ = \ {1\over2} \, ,
\fe
we find accordance with the field theory result~\eqref{Eqn:CJU1MOrb}.

Note that the Kaluza-Klein modes have minimal ${\rm U}(1)_{\rm M}$ charge $n/ 2$, which follows from the periodicity $\theta_3\sim \theta_3+{4\pi\over n}$ in the orbifold theory~\cite{Bergman:2012kr}. This holographic argument for the normalization of ${\rm U}(1)_{\rm M}$ is in agreement with the mesonic charge defined in field theory, given by~\eqref{chargeRelation1},~\eqref{chargeRelation2}, and~\eqref{chargeRelation3}, of the meson operator, given in \eqref{meson1}, \eqref{meson2}, and \eqref{meson3}.

Let us briefly remark on the dual supergravity gauge fields for the ${\rm U}(1)_{\rm I}^{(a)}$ instantonic symmetries. 
In~\cite{Bergman:2012kr}, it was argued that all but one of them arise from the reduction of the R-R three-form $C_{3}$ on the nontrivial two-cycles at the orbifold fixed point of ${\rm HS}^{4}/\mathbb{Z}_{n}$. The remaining one is argued to arise by a reduction akin to the infrared ${\rm U}(1)_{\rm I}$ of the Seiberg theories (see Section~\ref{sect:D0CJ}). We leave the matching of the instantonic flavor central charge to future work, in light of the subtleties already present in the Seiberg theories, as discussed in Section~\ref{sect:D0CJ}.

\subsection{Baryonic flavor central charges}
\label{sec:GravityBS}

The baryonic gauge fields are obtained from the reduction of the R-R three-form $C_3$ in ten-dimensional massive IIA on the internal two-cycles $\widetilde\Sigma_{a}$ in ${\rm HS}^{4}/\mathbb{Z}_{n}$ \cite{Bergman:2012kr}. Thus, in order to match the large $N$ baryonic flavor central charge matrix with the coupling matrix of the gauge fields in the holographic dual, we are required to analyze the appropriate (baryonic) two-cycles, and their intersection forms.

The orbifolded four-hemisphere, ${\rm HS}^{4}/\mathbb{Z}_{n}$, has an orbifold singularity at the north pole. Thus, the usual Kaluza-Klein reduction from IIA supergravity needs to be complemented by analyzing the twisted sectors of the string theory. We start by considering the covering space ${\rm S}^{4}/\bZ_{n}$ (before the orientifold projection), or more precisely a mirror pair of $\bC^2/\bZ_n$ singularities, \textit{before} the near horizon limit. We introduce two sets of vanishing anti-self-dual two-cycles $\sigma^{\rm N,S}_{i}$, with $i=1,\dots,n-1$, for the two copies of $\bC^2/\bZ_n$, corresponding to the twisted sector ground states of the orbifold CFT. Their intersection forms are given by the $A_{n-1}$ Cartan matrices,
\ie
(\, \sigma_i^{\rm N, S} \,,\,  \sigma_j^{\rm N, S} \,)  \ = \  \begin{pmatrix}
	2 \ & \ -1 \ & \ 0 \ & \ 0 \ &\cdots & \ 0 \ & \ 0 \ \\
	-1 \ & \ 2 \ & \ -1 \ & \ 0 \ &\cdots & \ 0 & \ 0 \ \\
	& & & \ \vdots \ &  &  \\
	0 \ & \ 0 \ &\cdots & \ 0 \ & \ -1 \ & \ 2 \ & \ -1 \ \\
	0 \ & \ 0 \ &\cdots & \ 0 \ & \ 0 \ & \ -1 \ & \  2 \\
\end{pmatrix}_{(n-1)\times (n-1)} \,
\fe
and
\ie
(\, \sigma_i^{\rm N} \,,\,  \sigma_j^{\rm S} \,)  \ = \ 0 \,.
\fe
As we shall explain below, the twisted sector closed string states that survive the orientifold projection give rise to six-dimensional gauge fields dual to the baryonic symmetries.

The orientifold action $\Omega I$, given by a composition of worldsheet parity $\Omega$ and the reflection ${I: x^{9} \ \to \ - x^{9}}$  (before the near-horizon limit), maps the vanishing cycles in the north and south copies of $\bC^2/\bZ_n$ as follows
\ie
\Omega I:~ \sigma^{\rm N}_i \ \leftrightarrow \ -\sigma^{\rm S}_{n-i} \,,
\fe 
for even $n=2k$ with $i\neq k$, and odd $n=2k+1$. For even $n=2k$, the middle cycles $\sigma^{\rm N,S}_k$ are allowed to transform in two different ways corresponding to with and without vector structure \cite{Berkooz:1996iz,Polchinski:1996ry,Intriligator:1997kq,Blum:1997fw}
\ie
{\rm vs:}\quad &\sigma^{\rm N}_k \ \leftrightarrow \ -\sigma^{\rm S}_k\,,
\\
{\rm nvs:}\quad &\sigma^{\rm N}_k \ \leftrightarrow \  \sigma^{\rm S}_k \,.
\fe

To make contact with the notation in reference~\cite{Bergman:2012kr}, we identify the cycles $\Sigma_i$ and $\widetilde\Sigma_i$ as  (again, we emphasize that this is prior to taking the near-horizon limit)\footnote{In~\cite{Bergman:2012kr}, upon Kaluza-Klein reduction of the R-R three-form $C_{3}$ from massive IIA, the different cycles $\Sigma_i$ and $\widetilde \Sigma_i$ give rise to six-dimensional ${\rm U}(1)$ gauge fields sourced by instantonic and baryonic ${\rm U}(1)$ global conserved currents, respectively.
}
\ie
&i<{n\over 2} \ : \ 
\Sigma_i \ = \  \sigma^{\rm N}_i+\sigma^{\rm S}_i \,,\qquad 
\widetilde\Sigma_i \ = \ \sigma^{\rm N}_i-\sigma^{\rm S}_{i}\,, 
\\
&i>{n\over 2} \ : \
\Sigma_i \ = \ - \sigma^{\rm N}_i-\sigma^{\rm S}_i \,, \qquad 
\widetilde\Sigma_i \ = \ \sigma^{\rm S}_i-\sigma^{\rm N}_{i}\,.
\fe
For even $n=2k$, there is an additional middle cycle, which we define in terms of $\sigma_{k}^{\rm N,S}$ as
\ie
\Sigma_k \ = \ \sigma^{\rm N}_k+\sigma^{\rm S}_k\,, \qquad 
\widetilde\Sigma_k \ = \ \sigma^{\rm N}_k-\sigma^{\rm S}_k\,.
\fe
Note that while there are D2-branes wrapping combinations of $\Sigma_i$ and sitting on the O8-plane, corresponding to (singular) instantons in the boundary theory, the baryonic cycles $\widetilde\Sigma_i$ only exist away from the O8-plane.

Then the six-dimensional baryonic gauge fields (coupled to the baryonic ${\rm U}(1)_{\rm B}$ currents in field theory),
are given by reducing the R-R three-form $C_{3}$ on linear combinations of the two-cycles $\widetilde\Sigma_i$ that are odd under the orientifold action $\Omega I$, that we call $\widetilde\Omega_a$,\footnote{Similarly, there are cycles $\Omega_a$ which are $\Omega I$-odd combinations of $\Sigma_i$. They give rise to six-dimensional gauge fields dual to instanton symmetries. However due to the mixing with other R-R and NS-NS fields from ten-dimensional supergravity as explained in Section~\ref{sect:D0CJ}, the intersection matrix of $\Omega_a$ will not directly produce the $C_J$ matrix for ${\rm U}(1)_{\rm I}$. We leave the complete analysis in this case for future.
}
\ie
B_a \ = \ \int_{\widetilde\Omega_a} C_3 \,.
\fe
Thus, the coupling matrix for the baryonic gauge fields are proportional to the intersection matrices $(\, \widetilde\Omega_a \,,\,  \widetilde \Omega_b \,)$ which we compute for each case of the orbifolds below.

\subsubsection*{Even orbifold with vector structure}
The odd two-cycles are
\ie
\widetilde\Omega_a \ = \ \widetilde \Sigma_a+\widetilde \Sigma_{n-a} \ = \  \left( \sigma^{\rm N}_a-\sigma^{\rm N}_{n-a}\right)-\left(\sigma^{\rm S}_{a}-\sigma^{\rm S}_{n-a}\right)  \,, \quad a = 1,\cdots,k-1 \,.
\fe
The intersection matrix is
\ie
& (\, \widetilde\Omega_a \,,\, \widetilde\Omega_b \,) \ = \ 4\times \begin{pmatrix}
	2 \ & \ -1 \ & \ 0 \ & \ 0 \ &\cdots & \ 0 \ & \ 0 \ \\
	-1 \ & \ 2 \ & \ -1 \ & \ 0 \ &\cdots & \ 0 & \ 0 \ \\
	& & & \ \vdots \ &  &  \\
	0 \ & \ 0 \ &\cdots & \ 0 \ & \ -1 \ & \ 2 \ & \ -1 \ \\
	0 \ & \ 0 \ &\cdots & \ 0 \ & \ 0 \ & \ -1 \ & \  2 \\
\end{pmatrix}_{(k-1)\times (k-1)} \, .
\label{IMvs}
\fe

\subsubsection*{Even orbifold without vector structure}
The odd two-cycles are
\ie
\widetilde\Omega_a \ = \ \widetilde \Sigma_a+\widetilde \Sigma_{n-a} \ = \  \left( \sigma^{\rm N}_a-\sigma^{\rm N}_{n-a}\right)-\left(\sigma^{\rm S}_{a}-\sigma^{\rm S}_{n-a}\right)  \,, \quad a = 1,\cdots,k-1 \,,
\fe
and 
\ie
\widetilde\Omega_k \ = \ \widetilde\Sigma_k \ = \  \sigma^{\rm N}_k-\sigma^{\rm S}_{k} \,.
\fe
Their intersection matrix is
\ie
& (\, \widetilde\Omega_a \,,\, \widetilde\Omega_b \,) \ = \ 4\times \begin{pmatrix}
	2 \ & \ -1 \ & \ 0 \ & \ 0 \ &\cdots & \ 0 \ & \ 0 \ \\
	-1 \ & \ 2 \ & \ -1 \ & \ 0 \ &\cdots & \ 0 & \ 0 \ \\
	& & & \ \vdots \ &  &  \\
	0 \ & \ 0 \ &\cdots & \ 0 \ & \ -1 \ & \ 2 \ & \ -1 \ \\
	0 \ & \ 0 \ &\cdots & \ 0 \ & \ 0 \ & \ -1 \ & \  2 \\
\end{pmatrix}_{(k-1)\times (k-1)}\,,
\\
& (\, \widetilde\Omega_a \,,\,\widetilde\Omega_{k} \,) \ = \ 0 \,, \quad (\, \widetilde\Omega_{k} \,,\,\widetilde\Omega_{k} \,) \ = \ 4 \,,
\label{IMnvs}
\fe
for $a,b=1,\cdots,k-1$.

\subsubsection*{Odd orbifold}
The odd cycles are
\ie
\widetilde\Omega_a \ = \ \widetilde \Sigma_a+\widetilde \Sigma_{n-a} \ = \  \left( \sigma^{\rm N}_a-\sigma^{\rm N}_{n-a}\right)-\left(\sigma^{\rm S}_{a}-\sigma^{\rm S}_{n-a}\right) \,, \quad a = 1,\cdots,k \,.
\fe
Their intersection matrix is
\ie
(\, \widetilde\Omega_a \,,\,  \widetilde\Omega_b \,) \ = \ 4\times \begin{pmatrix}
	2 \ & \ -1 \ & \ 0 \ & \ 0 \ &\cdots & \ 0 \ & \ 0 \ \\
	-1 \ & \ 2 \ & \ -1 \ & \ 0 \ &\cdots & \ 0 & \ 0 \ \\
	& & & \ \vdots \ &  &  \\
	0 \ & \ 0 \ &\cdots & \ 0 \ & \ -1 \ & \ 2 \ & \ -1 \ \\
	0 \ & \ 0 \ &\cdots & \ 0 \ & \ 0 \ & \ -1 \ & \  3 \\
\end{pmatrix}_{k\times k}\,.
\label{IModd}
\fe

The intersection matrices \eqref{IModd}, \eqref{IMnvs}, and \eqref{IMvs} are in agreement with the baryonic flavor central charge matrices \eqref{CJBmatrix}, up to an overall normalization that can be fixed.

\section{Conclusions}
\label{Sec:Conclusions}

In this paper, we study the central charges of a certain class of holographic five-dimensional superconformal field theories that have constructions in type I' string theory, from both the field theory and supergravity perspective. On the field theory side, we employ the formulae discovered by the present authors in previous work~\cite{Chang:2017cdx}, that relate particular deformations of the five-sphere partition function to the conformal and flavor central charges. In order to compare to the corresponding supergravity quantities, we take the large $N$ limit of the field theory, where it is argued that the instanton contributions are exponentially suppressed. Thus, we obtain exact large $N$ results purely based on the perturbative part of the partition function, and compute the large $N$ conformal and flavor central charges for our class of theories. The comparison of the large $N$ flavor central charges for the manifest infrared instantonic ${\rm {\rm U}(1)_{\rm I}}$ and hypermultiplet ${\rm SO(2N_{\rm f})}$ symmetries provides evidence for the flavor symmetry enhancement to $E_{{\rm N_{\rm f}}+1}$ in the ultraviolet. We further support our large $N$ results by juxtaposing the results from direct numerical integration, the saddle point method, and the analytic large $N$ formulae.

We then compare and explicitly match these large $N$ field theory results against their holographic duals. The central charges are related to certain couplings in the six-dimensional effective gravity dual that arise from the reduction of massive IIA supergravity. Although the corresponding ten-dimensional supergravity backgrounds have a curvature singularity in the internal manifold, by explicitly reducing the relevant terms in the ten-dimensional action to six dimensions, we obtain finite values for the effective six-dimensional couplings, that precisely match with the conformal and flavor central charges in the field theory.

The matching of the instantonic flavor central charges is left for future work. On the gravity side, as argued in Section~\ref{sect:D0CJ}, in order to reproduce the correct effective six-dimensional kinetic term from which the corresponding central charge can be extracted, we are required to take into account the reduction of all the fields in ten-dimensional supergravity. On the field theory side, we found only one independent instantonic flavor central charge to leading order in the large $N$ limit, \Ie, the flavor central charge matrix is rank-one. To capture the remaining (independent) central charges, we need to carry the matrix model analysis for the orbifold theories to further subleading order in $1/N$.

Finally, we are confident that a similar large $N$ analysis should provide evidence and checks for some proposed large $N$ dualities of five-dimensional theories~\cite{Bergman:2013aca}, as well as the recently discovered type IIB AdS$_{6}$ solutions~\cite{DHoker:2016ujz,DHoker:2016ysh}.\footnote{While AdS$_{6}$ solutions in massive IIA supergravity are rather rare~\cite{Passias:2012vp}, there is a variety of solutions in type IIB that contains a warped AdS$_{6}$ factor. Some early works in this direction include \cite{Lozano:2012au,Jeong:2013jfc}.
}

\section*{Acknowledgments}

We are grateful to Oren Bergman, Daniel L. Jafferis, Igor R. Klebanov, Silviu S. Pufu, and Diego Rodr{\'\i}guez-G{\'o}mez for helpful discussions and correspondence.  CC, YL, and YW thank the Aspen Center for Physics, MF thanks the Simons Summer Workshop, and YL thanks National Taiwan University for hospitality during the course of this work.   
 CC is supported in part by the U.S. Department of Energy grant DE-SC0009999. MF is supported by the David and Ellen Lee Postdoctoral Scholarship, YL is supported by the Sherman Fairchild Foundation, and both MF and YL by the U.S. Department of Energy, Office of Science, Office of High Energy Physics, under Award Number DE-SC0011632.  YW is supported in part by the US NSF under Grant No.~PHY-1620059 and by the Simons Foundation Grant No.~488653.  This work was partially performed at the Aspen Center for Physics, which is supported by National Science Foundation grant PHY-1607611.  

\appendix

\section{Triple Sine Function}
\label{App:triplesine}

In this Appendix, we introduce the triple sine function as well as its asymptotic limit, which is important when taking the large $N$ limit of the squashed five-sphere free energy. 

The multiple sine function is defined as
\ie
 S_N (z \mid \omega_1,\ldots, \omega_N) \ &= \  \Gamma_{N}(z \mid \omega_1, \ldots, \omega_N)^{-1} \ \Gamma_{N} (\omega_{\mathrm{tot}} - z \mid \omega_1, \ldots, \omega_N)^{(-1)^{N}} \, ,
\fe
where $\omega_{\rm tot} = \sum_{b=1}^{N} \omega_{j}$ and where $\Gamma_{N}$ is the multiplet gamma function defined as
\ie
\Gamma_{N}(z \mid \omega_1, \ldots, \omega_N) \ = \  & \exp \left[  \Psi_N\left( z \mid \omega_1, \ldots, \omega_N \right) \right]\,.
\fe
Here, we have introduced yet another special function
\ie
\Psi_N\left( z \mid \omega_1, \ldots, \omega_N \right)  \ &= \ \left. \frac{\diff}{\diff s}\right|_{s=0} \zeta_N \left( s,z \mid \omega_1, \ldots , \omega_N \right) \,,
\fe
where the multiple zeta-function is given as follows
\ie
\zeta_N \left( s,z \mid \omega_1, \ldots , \omega_N \right)  \ &= \  \sum_{m_1, \ldots, m_N = 0}^{\infty} \left( z+m_1 \omega_1 + \cdots m_N \omega_N \right)^{-s} \,,
\fe
where $\Real z>0$, $\Real s >N$ and $\omega_1, \ldots, \omega_N>0$. The function $\zeta_{N}$ is a meromorphic function with simple poles at $s=1, \ldots, N$.

\subsection{Asymptotics of the triple sine function}\label{AsymptTripSine}

In order to compute the large $N$ limit of the (squashed) five-sphere free energy, we require the asymptotic $\left| z \right| \rightarrow \infty$ expansion of $\log S_{3}\left( z \right)$ and we shall quickly mention how to compute those. In~\cite{Ruijsenaars:2000}, the author proved that there is an alternative definition of $\Psi_N \left( z \right)$ given as
\ie
 \Psi_N \left( z \mid  \omega_1, \ldots , \omega_N \right) \ &= \  \frac{(-1)^{N+1}}{N!} B_{N,N}(z) \log z +(-1)^{N} \sum_{k=0}^{N-1} \frac{B_{N,k}(0)z^{N-k}}{k! (N-k)!}\sum_{\ell=1}^{N-k} \frac{1}{\ell} + \\ 
 & +\sum_{k=N+1}^{M} \frac{(-1)^{k}}{k!}B_{N,k}\left( 0 \right) z^{N-k} (k-N-1)!+\mathcal{R}_{N,M} (z) \, , \label{Psiexp}
\fe
where $\Real z>0$, and $M\geq N$ is an arbitrary integer. $ \mathcal{R}_{N,M} (z)$ is some remainder, which was shown in~\cite{Ruijsenaars:2000} to behave as
\ie
 \mathcal{R}_{N,M} \left( z \right) \ \sim \ & \mathcal{O} \left( z^{N-M-1} \right) \, ,
\fe
in the asymptotic limit $|z|\rightarrow \infty$ as long as $\left|\arg z\right| < \pi$. Similarly it is straightforward to see that the third term in~\eqref{Psiexp} is of order $ \mathcal{O} \left( z^{-1} \right)$ in the asymptotic limit $|z|\rightarrow \infty$. Furthermore, we denoted by $B_{N,N}\left( z \mid \vec{\omega}\right)$ the generalized/multiple Bernoulli polynomials, which can be explicitly computed by expanding and solving
\ie
 \frac{t^{N} e^{z t}}{\prod^{N}_{b=1} \left( e^{\omega_b t} -1 \right)} \ &= \   \sum_{n=0}^{\infty}\frac{t^{n}}{n!} B_{N,n}\left( z \right)
\fe
order-by-order. For the case of interest in the present paper, \Ie~$N=3$, we have
\ie
B_{3,3}\left( z \mid \vec{\omega}\right) \ &= \  \frac{z^3}{\omega_1 \omega_2 \omega_3}-\frac{3 \omega_{\mathrm{tot}}}{2 \omega_1 \omega_2 \omega_3}z^2+\frac{\omega_{\mathrm{tot}}^2+\left(\omega_1 \omega_2+\omega_1 \omega_3 +\omega_2 \omega_3\right)}{2 \omega_1 \omega_2 \omega_3} z\\
&-\frac{\omega_{\mathrm{tot}}\left(\omega_1 \omega_2+\omega_1 \omega_3+\omega_2 \omega_3\right)}{4 \omega_1 \omega_2 \omega_3} \,.
\fe
Thus, it is now easy to compute~\eqref{Psiexp} in the asymptotic limit and equivalently for the triple sine function. 
Explicitly one obtains
\ie\label{S3Asym1}
& \log S_3\left( \ii z \mid \vec{\omega} \right) + \log S_3\left( -\ii z \mid \vec{\omega} \right)
\\
& \ \sim \
 -\frac{\pi}{3 \, \omega_1\omega_2\omega_3} |z|^{3} 
 +\frac{\pi \left( \omega_{\mathrm{tot}}^{2} + \omega_1 \omega_2+\omega_1 \omega_3+\omega_2 \omega_3 \right)}{6 \,\omega_1\omega_2\omega_3} |z| + {\cal O}(|z|^{2-M})\,,\\
\fe
as well as
\ie\label{S3Asym2}
 \log S_3\left( \ii z+ \tfrac{\omega_{\mathrm{tot}}}{2} \mid\vec{\omega}\right)
\ \sim \ & - \frac{\pi}{6 \, \omega_1\omega_2\omega_3} |z|^{3} - \frac{\pi \left( \omega^{2}_1+\omega^{2}_2 +\omega^{2}_3 \right)}{24 \, \omega_1\omega_2\omega_3} |z| + {\cal O}(|z|^{2-M})\,,
\fe
where $z \in \mathbb{R}$.  In other words, for arbitrary $M$, the contributions from the third term in \eqref{Psiexp} exactly cancel. Since the integer $M$ can be chosen arbitrarily large, the expansions \eqref{S3Asym1} and \eqref{S3Asym2} have no subleading power law corrections $|z|^{-n}$, for $n\ge 0$.

\section{Numerical evaluation of central charges}

\label{App:Numerics}

We present tabulated values for the round sphere free energy $-F_0$, the conformal central charge $C_T$, the mesonic flavor central charge $C_J^{{\rm SU}(2)_{\rm M}}$, and the exceptional flavor central charges $C_J^{G_{\rm f}}$ for $G_{\rm f} = E_1, E_2, \dotsc, E_8$, for Seiberg theories up to rank three.  Each quantity is extracted from the perturbative partition function \eqref{eqn:USpPartition}, computed via direct numerical integration and an {\it a priori} illegal saddle point approximation.  An important conclusion we draw is that the finite $N$ saddle point method actually produces good approximations for the integrals.

\begin{table}[H]
\centering
~\hspace{-.4in}
\begin{tabular}{|c|c||ccc|ccc|ccc|}
\hline
\multicolumn{2}{|c||}{\multirow{3}{*}{$-F_0$}}
& \multicolumn{9}{c|}{$N$}
\\
\hhline{~~---------}
\multicolumn{2}{|c||}{} & \multicolumn{3}{c|}{1} & \multicolumn{3}{c|}{2} & \multicolumn{3}{c|}{3}
\\
\hhline{~~---------}
\multicolumn{2}{|c||}{} & Integral & Saddle & Error & Integral & Saddle & Error & Integral & Saddle & Error
\\\hline\hline
\multirow{8}{*}{$G_{\rm f}$}
 & $ {E}_1 $ & $ 5.0967 $ & $ 5.2612 $ & $ 1.6 \% $ & $ 22.190 $ & $ 22.220 $ & $ 0.067 \% $ & $ 55.114 $ & $ 54.960 $ & $ -0.14 \% $ \\
 & $ {E}_2 $ & $ 6.1401 $ & $ 6.2817 $ & $ 1.1 \% $ & $ 25.425 $ & $ 25.398 $ & $ -0.052 \% $ & $ 61.896 $ & $ 61.645 $ & $ -0.20 \% $ \\
 & $ {E}_3 $ & $ 7.3949 $ & $ 7.5109 $ & $ 0.78 \% $ & $ 29.335 $ & $ 29.243 $ & $ -0.16 \% $ & $ 70.122 $ & $ 69.760 $ & $ -0.26 \% $ \\
 & $ {E}_4 $ & $ 8.9590 $ & $ 9.0441 $ & $ 0.47 \% $ & $ 34.233 $ & $ 34.061 $ & $ -0.25 \% $ & $ 80.430 $ & $ 79.965 $ & $ -0.29 \% $ \\
 & $ {E}_5 $ & $ 11.007 $ & $ 11.052 $ & $ 0.20 \% $ & $ 40.391 $ & $ 40.404 $ & $ 0.016 \% $ & $ 93.965 $ & $ 93.454 $ & $ -0.27 \% $ \\
 & $ {E}_6 $ & $ 13.898 $ & $ 13.886 $ & $ -0.041 \% $ & $ 49.543 $ & $ 49.413 $ & $ -0.13 \% $ & $ 113.63 $ & $ 112.70 $ & $ -0.41 \% $ \\
 & $ {E}_7 $ & $ 18.538 $ & $ 18.440 $ & $ -0.27 \% $ & $ 64.642 $ & $ 64.010 $ & $ -0.49 \% $ & $ 143.41 $ & $ 144.08 $ & $ 0.23 \% $ \\
 & $ {E}_8 $ & $ 28.473 $ & $ 28.215 $ & $ -0.46 \% $ & $ 96.712 $ & $ 95.699 $ & $ -0.53 \% $ & $ 214.24 $ & $ 212.68 $ & $ -0.36 \% $ \\
\hline
\end{tabular}
\caption{The perturbative values of the round sphere free energy $-F_0$ in the rank-one to rank-three Seiberg theories, computed by numerical integration and by the saddle point approximation.}
\label{Tab:FCT}
\end{table}

\begin{table}[H]
\centering
~\hspace{-.1in}
\begin{tabular}{|c|c||ccc|ccc|ccc|}
\hline
\multicolumn{2}{|c||}{\multirow{3}{*}{$C_T$}}
& \multicolumn{9}{c|}{$N$}
\\
\hhline{~~---------}
\multicolumn{2}{|c||}{} & \multicolumn{3}{c|}{1} & \multicolumn{3}{c|}{2} & \multicolumn{3}{c|}{3}
\\
\hhline{~~---------}
\multicolumn{2}{|c||}{} & Integral & Saddle & Error & Integral & Saddle & Error & Integral & Saddle & Error
\\\hline\hline
\multirow{8}{*}{$G_{\rm f}$}
 & $ {E}_1 $ & $ 333.39 $ & $ 365.53 $ & $ 4.6 \% $ & $ 1477.0 $ & $ 1529.8 $ & $ 1.8 \% $ & $ 3673.0 $ & $ 3741.8 $ & $ 0.93 \% $ \\
 & $ {E}_2 $ & $ 422.94 $ & $ 455.28 $ & $ 3.7 \% $ & $ 1737.1 $ & $ 1790.2 $ & $ 1.5 \% $ & $ 4197.1 $ & $ 4266.1 $ & $ 0.82 \% $ \\
 & $ {E}_3 $ & $ 529.78 $ & $ 562.40 $ & $ 3.0 \% $ & $ 2049.6 $ & $ 2102.9 $ & $ 1.3 \% $ & $ 4829.4 $ & $ 4898.5 $ & $ 0.71 \% $ \\
 & $ {E}_4 $ & $ 662.00 $ & $ 694.99 $ & $ 2.4 \% $ & $ 2438.4 $ & $ 2492.2 $ & $ 1.1 \% $ & $ 5619.4 $ & $ 5688.9 $ & $ 0.61 \% $ \\
 & $ {E}_5 $ & $ 834.00 $ & $ 867.48 $ & $ 2.0 \% $ & $ 2946.9 $ & $ 3001.4 $ & $ 0.92 \% $ & $ 6653.6 $ & $ 6727.3 $ & $ 0.55 \% $ \\
 & $ {E}_6 $ & $ 1075.1 $ & $ 1109.2 $ & $ 1.6 \% $ & $ 3663.9 $ & $ 3719.6 $ & $ 0.75 \% $ & $ 8126.6 $ & $ 8199.4 $ & $ 0.45 \% $ \\
 & $ {E}_7 $ & $ 1459.5 $ & $ 1494.3 $ & $ 1.2 \% $ & $ 4815.9 $ & $ 4873.8 $ & $ 0.60 \% $ & $ 10504. $ & $ 10580. $ & $ 0.36 \% $ \\
 & $ {E}_8 $ & $ 2274.4 $ & $ 2309.8 $ & $ 0.77 \% $ & $ 7287.9 $ & $ 7351.0 $ & $ 0.43 \% $ & $ 15651. $ & $ 15736. $ & $ 0.27 \% $ \\
\hline
\end{tabular}
\caption{The perturbative values of the conformal central charge $C_T$ in the rank-one to rank-three Seiberg theories, computed by numerical integration and by the saddle point approximation.}
\label{Tab:FCT}
\end{table}

\begin{table}[H]
\centering
\begin{tabular}{|c|c||ccc|ccc|}
\hline
\multicolumn{2}{|c||}{\multirow{3}{*}{$C_J^{{\rm SU}(2)_{\rm M}}$}}
& \multicolumn{6}{c|}{$N$}
\\
\hhline{~~------}
\multicolumn{2}{|c||}{} & \multicolumn{3}{c|}{2} & \multicolumn{3}{c|}{3}
\\
\hhline{~~------}
\multicolumn{2}{|c||}{} & Integral & Saddle & Error & Integral & Saddle & Error
\\\hline\hline
\multirow{8}{*}{$G_{\rm f}$}
 & $ {E}_1 $ & $ 66.277 $ & $ 66.218 $ & $ -0.045 \% $ & $ 218.62 $ & $ 218.15 $ & $ -0.11 \% $ \\
 & $ {E}_2 $ & $ 70.839 $ & $ 70.733 $ & $ -0.074 \% $ & $ 233.82 $ & $ 233.23 $ & $ -0.13 \% $ \\
 & $ {E}_3 $ & $ 76.440 $ & $ 76.268 $ & $ -0.11 \% $ & $ 252.45 $ & $ 251.77 $ & $ -0.13 \% $ \\
 & $ {E}_4 $ & $ 83.556 $ & $ 83.300 $ & $ -0.15 \% $ & $ 275.74 $ & $ 275.41 $ & $ -0.061 \% $ \\
 & $ {E}_5 $ & $ 93.064 $ & $ 92.708 $ & $ -0.19 \% $ & $ 307.17 $ & $ 307.14 $ & $ -0.0044 \% $ \\
 & $ {E}_6 $ & $ 106.81 $ & $ 106.34 $ & $ -0.22 \% $ & $ 354.67 $ & $ 353.31 $ & $ -0.19 \% $ \\
 & $ {E}_7 $ & $ 129.60 $ & $ 129.06 $ & $ -0.21 \% $ & $ 431.93 $ & $ 430.43 $ & $ -0.17 \% $ \\
 & $ {E}_8 $ & $ 180.73 $ & $ 180.35 $ & $ -0.11 \% $ & $ 605.43 $ & $ 604.63 $ & $ -0.067 \% $ \\
 \hline
\end{tabular}
\caption{The perturbative values of the mesonic flavor central charge $C_J^{{\rm SU}(2)_{\rm M}}$ in the rank-one to rank-three Seiberg theories, computed by numerical integration and by the saddle point approximation.}
\label{Tab:FCT}
\end{table}

\begin{table}[H]
\centering
~\hspace{-.25in}
\begin{tabular}{|c|c||ccc|ccc|ccc|}
\hline
\multicolumn{2}{|c||}{\multirow{3}{*}{$C_J^{{\rm SO}(2N_{\rm f})}$}}
& \multicolumn{9}{c|}{$N$}
\\
\hhline{~~---------}
\multicolumn{2}{|c||}{} & \multicolumn{3}{c|}{1} & \multicolumn{3}{c|}{2} & \multicolumn{3}{c|}{3}
\\
\hhline{~~---------}
\multicolumn{2}{|c||}{} & Integral & Saddle & Error & Integral & Saddle & Error & Integral & Saddle & Error
\\\hline\hline
\multirow{7}{*}{$G_{\rm f}$}
 & $ {E}_2 $ & $ 21.638 $ & $ 22.275 $ & $ 1.5 \% $ & $ 51.739 $ & $ 52.559 $ & $ 0.79 \% $ & $ 88.687 $ & $ 89.681 $ & $ 0.56 \% $ \\
 & $ {E}_3 $ & $ 23.700 $ & $ 24.400 $ & $ 1.5 \% $ & $ 56.609 $ & $ 57.504 $ & $ 0.78 \% $ & $ 96.877 $ & $ 97.981 $ & $ 0.57 \% $ \\
 & $ {E}_4 $ & $ 26.413 $ & $ 27.158 $ & $ 1.4 \% $ & $ 62.941 $ & $ 63.886 $ & $ 0.75 \% $ & $ 107.22 $ & $ 108.66 $ & $ 0.67 \% $ \\
 & $ {E}_5 $ & $ 30.131 $ & $ 30.895 $ & $ 1.3 \% $ & $ 71.551 $ & $ 72.512 $ & $ 0.67 \% $ & $ 121.22 $ & $ 123.07 $ & $ 0.76 \% $ \\
 & $ {E}_6 $ & $ 35.587 $ & $ 36.324 $ & $ 1.0 \% $ & $ 84.126 $ & $ 85.045 $ & $ 0.54 \% $ & $ 142.94 $ & $ 144.01 $ & $ 0.37 \% $ \\
 & $ {E}_7 $ & $ 44.657 $ & $ 45.256 $ & $ 0.67 \% $ & $ 105.01 $ & $ 105.76 $ & $ 0.35 \% $ & $ 177.82 $ & $ 178.69 $ & $ 0.24 \% $ \\
 & $ {E}_8 $ & $ 64.752 $ & $ 64.766 $ & $ 0.011 \% $ & $ 151.41 $ & $ 151.47 $ & $ 0.019 \% $ & $ 255.52 $ & $ 255.60 $ & $ 0.015 \% $ \\
 \hline
\end{tabular}
\caption{The perturbative values of the flavor central charge $C_J^{G_{\rm f}}$ in the rank-one to rank-three Seiberg theories, obtained using the hypermultiplet masses, computed by numerical integration and by the saddle point approximation.}
\label{Tab:CJSO}
\end{table}

\begin{table}[H]
\centering
~\hspace{-.3in}
\begin{tabular}{|c|c||ccc|ccc|ccc|}
\hline
\multicolumn{2}{|c||}{\multirow{3}{*}{$\displaystyle C_J^{{\rm U}(1)_{\rm I}} \over \displaystyle I_{\mathfrak{u}(1)_{\rm I} \hookrightarrow \mathfrak{g}_{\rm f}}$}}
& \multicolumn{9}{c|}{$N$}
\\
\hhline{~~---------}
\multicolumn{2}{|c||}{} & \multicolumn{3}{c|}{1} & \multicolumn{3}{c|}{2} & \multicolumn{3}{c|}{3}
\\
\hhline{~~---------}
\multicolumn{2}{|c||}{} & Integral & Saddle & Error & Integral & Saddle & Error & Integral & Saddle & Error
\\\hline\hline
\multirow{8}{*}{$G_{\rm f}$}
 & $ {E}_1 $ & $ 18.409 $ & $ 17.966 $ & $ -1.2 \% $ & $ 45.661 $ & $ 45.919 $ & $ 0.28 \% $ & $ 79.666 $ & $ 80.481 $ & $ 0.51 \% $ \\
 & $ {E}_2 $ & $ 20.582 $ & $ 19.956 $ & $ -1.5 \% $ & $ 50.318 $ & $ 50.370 $ & $ 0.051 \% $ & $ 87.144 $ & $ 87.726 $ & $ 0.33 \% $ \\
 & $ {E}_3 $ & $ 23.120 $ & $ 22.342 $ & $ -1.7 \% $ & $ 55.843 $ & $ 55.722 $ & $ -0.11 \% $ & $ 96.095 $ & $ 96.490 $ & $ 0.21 \% $ \\
 & $ {E}_4 $ & $ 26.190 $ & $ 25.307 $ & $ -1.7 \% $ & $ 62.655 $ & $ 62.411 $ & $ -0.19 \% $ & $ 107.25 $ & $ 107.51 $ & $ 0.12 \% $ \\
 & $ {E}_5 $ & $ 30.128 $ & $ 29.192 $ & $ -1.6 \% $ & $ 71.546 $ & $ 71.243 $ & $ -0.21 \% $ & $ 121.96 $ & $ 122.14 $ & $ 0.075 \% $ \\
 & $ {E}_6 $ & $ 35.664 $ & $ 34.722 $ & $ -1.3 \% $ & $ 84.218 $ & $ 83.907 $ & $ -0.18 \% $ & $ 143.05 $ & $ 143.22 $ & $ 0.058 \% $ \\
 & $ {E}_7 $ & $ 44.707 $ & $ 43.783 $ & $ -1.0 \% $ & $ 105.07 $ & $ 104.78 $ & $ -0.14 \% $ & $ 177.89 $ & $ 178.07 $ & $ 0.051 \% $ \\
 & $ {E}_8 $ & $ 64.756 $ & $ 63.825 $ & $ -0.72 \% $ & $ 151.42 $ & $ 151.12 $ & $ -0.10 \% $ & $ 255.42 $ & $ 255.64 $ & $ 0.043 \% $ \\
 \hline
\end{tabular}
\caption{The perturbative values of the flavor central charge $C_J^{G_{\rm f}}$ in the rank-one to rank-three Seiberg theories, obtained using the instanton particle mass, computed by numerical integration and by the saddle point approximation.}
\label{Tab:CJSO}
\end{table}

\bibliography{refs} 
\bibliographystyle{JHEP}

\end{document}